\documentclass[12pt]{article}
\usepackage{amsmath}
\usepackage{amsthm,hyperref}
\usepackage{amssymb,amsfonts}
\usepackage[mathscr]{eucal}
\usepackage{graphicx}
\usepackage{color}
\usepackage{enumerate}
\usepackage[round]{natbib}

\usepackage{natbib}
\usepackage{authblk}
\usepackage{setspace} 
\usepackage{float}

\def\argmax{\mathop{\rm arg\, max}}
\def\argmin{\mathop{\rm arg\, min}}

\newcommand{\bel}{\begin{eqnarray}\label}
\newcommand{\eel}{\end{eqnarray}}
\newcommand{\bes}{\begin{eqnarray*}}
\newcommand{\ees}{\end{eqnarray*}}
\newcommand{\bei}{\begin{itemize}}
\newcommand{\beiftnt}{\begin{itemize}\footnotesize}
\newcommand{\eei}{\end{itemize}}

\def\benu{\begin{enumerate}}
\def\eenu{\end{enumerate}}

\def\argmax{\mathop{\rm arg\, max}}
\def\argmin{\mathop{\rm arg\, min}}
\def\real{{\mathbb{R}}}
\def\R{{\real}}

\def\E{{\mathbb{E}}}

\def\complex{\mathop{{\rm I}\kern-.58em\hbox{\rm C}}\nolimits}

\def\Var{\hbox{\rm Var}}

\def\mathbold{\boldsymbol} 

\def\ba{\mathbold{a}}\def\ahat{\widehat{a}}

\def\bb{\mathbold{b}}\def\bhat{\widehat{b}}

\def\chat{\widehat{c}}

\def\chat{\widehat{c}}

\def\Gtil{{\widetilde G}}

\def\mhat{\widehat{m}}

\def\Phat{\widehat{P}}

\def\Vhat{\widehat{V}}\def\Vtil{{\widetilde V}}

\def\ytil{\widetilde{y}}

\def\Ztil{{\widetilde Z}}

\def\bfeta{\mathbold{\eta}}\def\etahat{\widehat{\eta}}
\def\etatil{\widetilde{\eta}}

\def\btheta{\mathbold{\theta}}

\def\lambdahat{\widehat{\lambda}}
\def\lambdatil{\widetilde{\lambda}}

\def\muhat{\widehat{\mu}}
\def\mutil{\widetilde{\mu}}

\def\sigmahat{\widehat{\sigma}}

\def\0{\mathbold{0}}
\def\1{\mathbold{1}}

\newtheorem{theorem}{Theorem}
\newtheorem{lemma}{Lemma}

\newtheorem{corollary}{Corollary}

\numberwithin{theorem}{section}
\numberwithin{lemma}{section}
\numberwithin{corollary}{section}

\theoremstyle{definition}

\newcommand{\thr}{m_{r,c}}
\newcommand{\net}{\Theta_{r,c}}
\newcommand{\nett}{\Theta}

\newcommand{\tr}{\hbox{\rm tr}}
\newcommand{\lambdaa}{\lambda_{A}}
\newcommand{\lambdab}{\lambda_{B}}

\newcommand{\lambdaahat}{\lambdahat_A}
\newcommand{\lambdabhat}{\lambdahat_B}
\newcommand{\lambdaatil}{\lambdatil_A}
\newcommand{\lambdabtil}{\lambdatil_B}
\newcommand{\Za}{Z_A}
\newcommand{\Zb}{Z_B}
\newcommand{\Vinv}{\Sigma^{-1}}
\newcommand{\Minv}{M^{-1}}
\renewcommand{\Vhat}{\hat{\Sigma}}
\newcommand{\Vhatinv}{\Vhat^{-1}}
\newcommand{\mhatls}{\mhat^{\text{LS}}}
\newcommand{\ahatls}{\ahat^{\text{LS}}}
\newcommand{\bhatls}{\bhat^{\text{LS}}}
\newcommand{\ahatbls}{\widehat{\bold{a}}^{\text{LS}}}
\newcommand{\bhatbls}{\widehat{\bold{b}}^{\text{LS}}}
\newcommand{\Zinv}{Z^{\dagger}}

\newcommand{\half}{\frac{1}{2}}
\newcommand{\rcinv}{\frac{1}{rc}}

\newcommand{\Gdot}{\dot{G}}
\newcommand{\alphab}{\ensuremath{\boldsymbol{\alpha}}}
\newcommand{\betab}{\ensuremath{\boldsymbol{\beta}}}
\newcommand{\etab}{\ensuremath{\boldsymbol{\eta}}}
\newcommand{\one}{\ensuremath{\boldsymbol{1}}}
\newcommand{\thetab}{\ensuremath{\boldsymbol{\theta}}}
\newcommand{\Y}{\ensuremath{\boldsymbol{Y}}}
\newcommand{\y}{\ensuremath{\boldsymbol{y}}}
\newcommand{\x}{\ensuremath{\boldsymbol{x}}}
\newcommand{\etahatb}{\ensuremath{\boldsymbol{\etahat}}}
\newcommand{\etahatsb}{\ensuremath{\boldsymbol{\etahat}^{\sf\,S}}}

\newcommand{\ex}{\ensuremath{\mathbb{E}}}
\newcommand{\V}{\ensuremath{\Sigma}}

\newcommand{\URE}{\ensuremath{\mathsf{\widehat{URE}}}}
\newcommand{\UREQ}{\ensuremath{\mathsf{\widehat{URE}^{\sf Q}}}}

\newcommand{\etatb}{\ensuremath{\widetilde{\etab}}}
\newcommand{\widesim}[2][1.5]{
	\mathrel{\overset{#2}{\scalebox{#1}[1]{$\sim$}}}
}
\newcommand{\etatiltil}{\widetilde{\etab}_{\sf c}}
\renewcommand{\Ztil}{\ensuremath{Z_{\sf c}}}
\renewcommand{\etatil}{\etab_{\sf c}}
\newcommand{\hetatil}{\hat{\etab}_{\sf c}}
\newcommand{\topp}{\mathsf{T}}

\newcount\Comments  
\definecolor{darkgreen}{rgb}{0,0.5,0}
\definecolor{purple}{rgb}{1,0,1}
\newcommand{\kibitz}[2]{\ifnum\Comments=1\textcolor{#1}{#2}\fi}

\title{\sc Empirical Bayes Estimates for a 2-Way Cross-Classified Additive Model}
\date{}
\author[1]{Lawrence D. Brown} 
\author[2]{Gourab Mukherjee} 
\author[1,3]{Asaf Weinstein}
\affil[1]{University of Pennsylvania}
\affil[2]{University of Southern California}
\affil[3]{Stanford University}

\begin{document}

\maketitle

\begin{abstract}
	We develop an empirical Bayes procedure for estimating the cell means in an unbalanced, two-way additive model with fixed effects. 
	We employ a hierarchical model, which reflects exchangeability of the effects within treatment and within block but not necessarily between them, as suggested before by \citet{lindley1972bayes}. 
	The hyperparameters of this hierarchical model, instead of considered fixed, are to be substituted with data-dependent values in such a way that the point risk of the empirical Bayes estimator is small. 
	Our method chooses the hyperparameters by minimizing an unbiased risk estimate and is shown to be asymptotically optimal for the estimation problem defined above. 
	The usual empirical Best Linear Unbiased Predictor (BLUP) is shown to be substantially different from the proposed method in the unbalanced case and therefore performs sub-optimally. 
	Our estimator is implemented through a computationally tractable algorithm that is scalable to work under large designs. 
	The case of missing cell observations is treated as well. 
	We demonstrate the advantages of our method over the BLUP estimator through simulations and in a real data example, where we estimate average nitrate levels in water sources based on their locations and the time of the day.
\end{abstract}

\noindent {\small{\it Some key words}: {Shrinkage estimation; Empirical Bayes; Two-way ANOVA;  Oracle Optimality; Stein's unbiased risk estimate (SURE); Empirical BLUP.}}

\section{Introduction} \label{sec:intro}
Multilevel cross-classified models are pervasive in statistics, with applications ranging from detecting sources of variability in medical research \citep{goldstein2002multilevel} to understanding micro-macro linkages in social studies \citep{mason1983contextual,zaccarin2002multilevel}. 
These models offer a natural and flexible approach to specify meaningful latent structures and, importantly, a systematic way to use all information  for simultaneously analyzing the effects of more than one factor \citep{rasb1994efficient}. 
Hierarchical cross-classified models have  classically been used to decompose the total variability of the response into individual sources and for prediction in random-effects models. 
Nevertheless, ever since the appearance of the James-Stein estimator \citep{james1961estimation} and its Bayesian interpretation \citep{stein1962confidence,lindley1962discussion}, the usefulness of such models in estimation problems involving multiple {\it nonrandom} effects has been well recognized. 

Hierarchical models have been used to facilitate shrinkage estimators in linear regression models since the early 1970s \citep{efron1972empirical}. 
In both theoretical and more applied work, various authors have employed hierarchical models to produce estimators that shrink {\it towards} a subspace \citep[e.g.,][]{sclove1968improved, oman1982shrinking, jiang2011best, tan2014steinized} or {\it within} a subspace \citep[e.g.,][]{lindley1972bayes, rolph1976choosing, kou2015optimal}; see Section 2 of the last reference for a discussion on the difference between the two types of resulting estimators. 
Cross-classified additive models are in a sense the most immediate extension of Stein's canonical example. 
Specifically, unlike in a general linear model, the symmetries of within-batch effects can be regarded as a-priori information, which suggest the use of exchangeable priors, such as those proposed by \citet{lindley1972bayes} and \citet{,efron1973stein}. 
In the case of balanced design, the properties of resulting shrinkage estimators are by now well understood and have a close relationship to the James-Stein estimator. 
Indeed, when all cell counts are equal, multiple one-way, homoscedastic estimation problems emerge; for these the James-Stein estimator has optimality properties under many criteria. 
But in the unbalanced case, the problems of estimating the effects corresponding to different batches are intertwined due to lack of orthogonality in the design matrix; hence, the situation in the case of unbalanced design is substantially different. 


This paper deals with empirical Bayes (EB) estimation of the cell means in the two-way fixed effects additive model with unbalanced design. 
We consider a family of Bayes estimators resulting from a normal hierarchical model, which reflects within-batch exchangeability and is indexed by a set of hyper-parameters that govern the prior. 
Any corresponding estimator that substitutes {\it data-dependent} values for the hyper-parameters is referred to as an empirical Bayes estimator. 
We propose an empirical Bayes procedure that is asymptotically optimal for the estimation of the cell means under squared loss. 
In our asymptotic analysis, the number of row and column levels tends to infinity. 
Importantly, the so-called empirical BLUP (Best Linear Unbiased Predictors) estimators, using the usual maximum-likelihood approach in estimating the hyperparameters, are shown to perform sub-optimally in the unbalanced case. 
Instead of using the maximum-likelihood criterion, we choose the values for the hyper-parameters by minimizing an unbiased estimate of the risk (URE), which leads to estimates that are different in an essential way. 
The proposed approach is appealing in the fixed effects case, because it uses a criterion directly related to the risk instead of using the likelihood under the postulated hierarchical model. 


Using the URE criterion to calibrate tuning parameters has been proposed in many previous works and in a broad range of parametric and nonparametric estimation problems \citep[][to name a few]{li1986asymptotic, ghosh1987sequential, donoho1995wavelet, johnstone2004needles, candes2013unbiased}. 
Recently, \citet{xie2012sure} employed URE minimization to construct alternative empirical Bayes estimators to the usual ones in the Gaussian mean problem with known {\it heteroscedastic} variances and showed that it produces asymptotically uniformly better estimates. 
Our work can be viewed as a generalization of \citet{xie2012sure} from the one-way unbalanced layout to the two-way unbalanced layout. 

The two-way unbalanced problem presents various new challenges. The basis for the difference, of course, lies in the facts that the two-way case imposes structure on the mean vector, which is nontrivial to handle due to missingness and imbalance in the design. 
Some of the implications are that the analysis of the performance of EB methods is substantially more involved than in the one-way scenario; in addition, the implementation of the URE estimator, which is trivial in the one-way scenario, becomes a cause of concern, especially with a growing number of factor levels. 
We offer an implementation of the corresponding URE estimate that in the all-cells-filled case has comparable computational performance to that of the standard empirical BLUP in the popular R package \texttt{lme4} of \citet{bates2010lme4}. Our theoretical analysis of the two-way case differs in fundamental aspects from the optimality proof techniques usually used in the one-way normal mean estimation problem. 
To tackle the difficulties encountered in the two-way problem, where computations involving matrices are generally unavoidable, we developed a flexible approach for proving asymptotic optimality based on efficient pointwise risk estimation; this essentially reduces our task to controlling the moments of Gaussian quadratic forms.    


We would also like to point out that the current work is different from the recent extensions of \citet{kou2015optimal} of the URE approach to the general Gaussian linear model. 
While the setup considered in that paper formally includes our setup as a special case, their results have limited implications for additive cross-classified models; 
for example, the covariance matrix used in their second level of the hierarchy is not general enough to accommodate the within-batch exchangeable structure we employ and is instead governed by a single hyper-parameter. 
Moreover, their asymptotic results require keeping the dimension of the linear subspace fixed, whereas the number of factor levels is increasing in our setup. 

\textbf{Organization of the paper.}
In Section~\ref{sec:bayes} we describe our estimation setup for the simplest case when there are no missing observations. 
In Section~\ref{sec:missing} we introduce the more general model, which allows missing observations, and describe a unified framework for estimation across all scenarios -- missing or non-missing. 
In Section~\ref{sec:theory} we show that our proposed estimation methodology is asymptotically optimal and is capable of recovering the directions and magnitude for optimal shrinkage; this is established through the notion of {oracle} optimality. 
Section~\ref{sec:bal} is devoted to the special case of a balanced design. 
After describing the computation details in Section~\ref{sec:computation}, we report the results from extensive numerical experiments 
in Section~\ref{sec:simulation}. 
Lastly, in Section~\ref{sec:real.data} we demonstrate the applicability of our proposed method on a real-world problem concerning the estimation of the average nitrate levels in water sources based on location and time of day.

\section{Model Setup and Estimation Methods} \label{sec:bayes}
\subsection{Basic Model and Estimation Setup}
\textbf{Additive model with all cells filled.} 
Consider the following basic two-way cross-classified additive model with fixed effects:
\begin{align}\label{eq:model-twoway}
\begin{split}
& y_{ij} = \eta_{ij} + \epsilon_{ij}, \quad \quad  1\leq i \leq r \text{ and } 1\leq j \leq c~, \\
 \text{ where} \quad & \eta_{ij} = \mu + \alpha_i + \beta_j \quad \text{ and } \quad \epsilon_{ij} \sim N(0,\sigma^2 K^{-1}_{ij}). 
\end{split}
\end{align}
$K_{ij}$ is the number of observations, or the {\it count} in the $(i,j)^{\text{th}}$ cell; $\sigma^2>0$ is assumed to be known; and $\epsilon_{ij}$ are independent Gaussian noise terms.  
Model~\eqref{eq:model-twoway} is over-parametrized, hence the parameters  $\mu, \alphab=(\alpha_1,...,\alpha_r)^{\topp}, \betab=(\beta_1, ...,\beta_c)^{\topp}$ are not identifiable without imposing further side conditions; however, the vector of cell means $\etab=(\eta_{11}, \eta_{12}, ..., \eta_{rc})^{\topp}$  is always identifiable. 
Our goal is to estimate $\etab$ under the sum-of-squares loss
\bel{eq:loss}
L_{r,c}(\etab, \etahatb) = \frac{1}{rc} \, \| \etahatb - \etab \|^2 = \frac{1}{rc}  \sum_{i=1}^r \sum_{j=1}^c (\etahat_{ij} - \eta_{ij})^2. 
\eel
In model \eqref{eq:model-twoway} the unknown quantities $\alpha_i$ and $\beta_j$ will be referred to as the $i$-th ``row'' (or ``treatment")  and the $j$-th ``column'' (or ``block") effects, respectively. 
In the {\it all-cells-filled} model, $K_{ij}\geq 1$ for $1\leq i \leq r \text{ and } 1\leq j \leq c$; 
the more general model, which allows some empty cells, is presented in Section~\ref{sec:missing}. 
We would like to emphasize the focus in this section on the loss \eqref{eq:loss} rather than the weighted quadratic loss 
\bes
L_{r,c}^{\text{wgt}}(\etab, \etahatb) = \frac{1}{rc}  \sum_{i=1}^r \sum_{j=1}^c K_{ij} (\etahat_{ij} - \eta_{ij})^2~,
\ees
which is sometimes called the ``prediction" loss, and under which asymptotically optimal estimation has been investigated before \citep{dicker2013optimal}. 
Nevertheless, in later sections results are presented for a general quadratic loss, which includes the weighted loss as a special case. 

\smallskip
\textbf{Shrinkage estimators for the two-way model.} 
The usual estimator of $\etab$ is the weighted least squares (WLS) estimator, which is also maximum-likelihood under \eqref{eq:model-twoway}. 
The WLS estimator is unbiased and minimax but can be substantially improved on in terms of quadratic loss by shrinkage estimators, particularly when $r,c \to \infty$ \citep{draper1979ridge}. 
Note that through out this paper we represent vectors in bold and matrices by capital letters.
As the starting point for the shrinkage estimators proposed in this paper, we consider a family of Bayes estimators with respect to a conjugate prior on $(\alphab,\betab)$
\begin{align*}
\alpha_1,\ldots,\alpha_r  \widesim{\text{i.i.d.}} N(0, \sigma^2_A) \quad \text{ and } \quad \beta_1, \ldots,\beta_c  \widesim{\text{i.i.d.}} N(0, \sigma^2_B)~, 
\end{align*}
where $\sigma^2_A, \sigma^2_B$ are hyper-parameters. 
This prior extends the conjugate normal prior in the one-way case and was proposed by \citet{lindley1972bayes} to reflect exchangeability within rows and columns separately. 
In vector form, the two-level hierarchical model is:
\begin{align}
\begin{split}\label{eq:model-bayes}
\text{Level 1:} & \quad \y|\etab \sim N_{p}(\etab, \sigma^2 M) \ \ \ \ \ \ \etab =   \one\mu  + Z\thetab \ \ \ \  \thetab^{\topp}=(\alphab^{\topp},\betab^{\topp})\\
\text{Level 2:} & \quad \thetab \sim N_q(0, \sigma^2\Lambda\Lambda^{\topp})~,
\end{split}
\end{align}
where $M = \text{diag}(K_{11}^{-1}, K_{12}^{-1},...,K_{rc}^{-1})$ is an $rc \times rc$ matrix
and $Z = [\Za\ \Zb]$ with $\Za = I_r \otimes 1_c$ and $\Zb = 1_r \otimes I_c$. 
The $(r+c)\times (r+c)$ matrix 
\bes
\begin{aligned}
 \Lambda = 
\begin{bmatrix}
\sqrt{\lambdaa}\; I_r & 0 \\
0 & \sqrt{\lambdab} \;I_c
\end{bmatrix}\;
\end{aligned}
\ees
is written in terms of the relative variance components $\lambda_A = {\sigma^2_A/\sigma^2}$ and $\lambda_B = {\sigma^2_B/\sigma^2}$. 
Henceforth, for notational simplicity, the dependence of $\Lambda$ on the model hyper-parameters will be kept implicit. As shown in Lemma~\ref{lem.misc.1} of the supplementary materials, the marginal variance of $\y$ in \eqref{eq:bayes} is given by $\sigma^2\, \Sigma$ where
\begin{align}\label{eq:V-compound} 
\V = Z\Lambda \Lambda^{\topp} Z^{\topp} + M= \lambdaa \Za \Za^{\topp} + \lambdab \Zb \Zb^{\topp} + M.
\end{align}

At this point a comment is in order regarding shrinkage estimators for the general homoscedastic linear model. 
Note that model \eqref{eq:model-twoway} could be written for individual, homoscedastic observations (with an additional subscript $k$) instead of for the cell averages. 
With the corresponding design matrix, the two-way additive model is therefore a special case of the homoscedastic Gaussian linear model, $\y \sim N_n(X\boldsymbol{\gamma}, \sigma^2 I)$, where $X\in \R^{n\times p}$ a known matrix and $\boldsymbol{\gamma} \in \R^p$ is the unknown parameter. 
Thus, the various Stein-type shrinkage methods that have been proposed for estimating $\boldsymbol{\gamma}$ can also be applied to our problem. 
Specifically, a popular approach is to reduce the problem of estimating $\boldsymbol{\gamma}$ to the problem of estimating the mean of a $p$-dimensional {\it heteroscedastic} normal vector with known variances \citep[see, e.g., ][Section 2.9]{johnstone2011gaussian} by applying orthogonal transformations to the parameter $\boldsymbol{\gamma}$ and data $\y$. 
Thereafter, Stein-type shrinkage estimators can be constructed as empirical Bayes rules by putting a prior which is either i.i.d. on the transformed coordinates or i.i.d. on the original coordinates of the parameter \citep[][referred to priors of the first type as {\it proportional} priors and to those of the second kind as {\it constant} priors]{rolph1976choosing}.
In the case of factorial designs, however, neither of these choices is very sensible, because they do not capture the (within-batch) symmetries of cross-classified models. 
Hence, procedures relying on models that take exchangeability into account can potentially achieve a significant and meaningful reduction in estimation risk. The estimation methodology we develop here incorporates the exchangeable structure of \eqref{eq:model-bayes}.


\smallskip
\textbf{Empirical Bayes estimators.} 
The following is a standard result and is proved in Section~\ref{append.sec2} of the supplementary materials.  

\begin{lemma}
	For any fixed $\mu \in \R$ and non-negative $\lambdaa$, $\lambdab$  the Bayes estimate of $\etab$ in \eqref{eq:model-bayes} is given by:
	\begin{align}\label{eq:bayes}
		E[\etab|\y] = \y - M \Vinv (\y -  \one\mu) \;,	
	\end{align}
	where the hyper-parameters $\lambdaa$, $\lambdab$ are involved in $\V$ through $\Lambda$. 
\end{lemma}

Instead of fixing the values of $\mu, \lambdaa, \lambdab$ in advance, we may now return to model \eqref{eq:model-twoway} and consider the parametric family of estimators
\bel{eq:bayes-family}
\mathcal{S}[\tau]=\Big\{ \etahatsb(\mu, \lambdaa, \lambdab)=\y - M \Vinv (\y -  \one\mu): \mu \in [\hat{a}_{\tau}(\y), \hat{b}_{\tau}(\y)],\ \lambdaa \geq 0,\ \lambdab \geq 0 \Big\}. 
\eel
Above, $\mu$ is restricted to lie within $\hat{a}_{\tau}(\y)=\texttt{quantile}\{y_{ij}: 1 \leq i \leq r, 1 \leq j \leq c; \, \tau/2 \}$ and $\hat{b}_{\tau}(\y)=\texttt{quantile}\{y_{ij}: 1 \leq i \leq r, 1 \leq j \leq c; 1- \tau/2 \}$, the $\tau/2$ and $(1-\tau/2)$ quantiles of the observations. 
The constraint on the location hyper-parameter $\mu$ is imposed for technical reasons but is moderate enough to be well justified. 
Indeed, an estimator that shrinks toward a point that lies near the periphery or outside the range of the data is at the risk of being non-robust and seems to be an undesirable choice for a Bayes estimator correponding to \eqref{eq:model-bayes}, which models $\alphab$ and $\betab$ as having zero means. 
In practice $\tau$ may be taken to be 1\% or 5\%. 

An empirical Bayes estimator is obtained by selecting for each observed $\y$ a (possibly different) candidate from the family $\mathcal{S}[\tau]$ as an estimate for $\etab$; equivalently, an empirical Bayes estimator is any estimator that plugs data-dependent values $\muhat, \lambdaahat, \lambdabhat$ into \eqref{eq:bayes}, with the restriction that $\muhat$ is in the allowable range. 
In the next section, we propose a specific criterion for estimating the hyperparameters. 

\subsection{Estimation Methods} 

The usual empirical Bayes estimators are derived relying on hierarchical model \eqref{eq:model-bayes}. The fixed effect $\mu$ and the relative variance components $\lambdaa$ and $\lambdab$ are treated as unknown fixed parameters to be estimated based on the marginal distribution of $\y$ and substituted into \eqref{eq:bayes}. For any set of estimates substituted for $\lambdaa$ and $\lambdab$, the general mean $\mu$ is customarily estimated by generalized least squares, producing an empirical version of what is known as the Best Linear Unbiased Predictors (BLUP). 
There is extensive literature on the estimation of the variance components  (see chapters 5 and 6 of \citealp{searle2009variance}), with the main methods being maximum-likelihood (ML), restricted maximum-likelihood (REML), and the ANOVA methods (Method-of-Moments), including the three original ANOVA methods of Henderson \citep{henderson1984anova}. 
Here we concentrate on the commonly used maximum-likelihood estimates, which are implemented in the popular \texttt{R} package \texttt{lme4} \citep{lme4}. 
If $\mathcal{L} (\mu, \lambdaa, \lambdab; \y)$ denotes the marginal likelihood of $\y$ according to \eqref{eq:model-bayes}, then the maximum-likelihood (ML) estimates are
\bel{eq:ebmle-hyper}
(\muhat^{\rm ML}, \lambdaahat^{\rm ML}, \lambdabhat^{\rm ML}) = \argmax_{\mu \in [\hat{a}_{\tau},\hat{b}_{\tau}], \lambdaa\geq 0, \lambdab\geq 0} \; \mathcal{L} ( \mu, \lambdaa, \lambdab; \y). 
\eel
The corresponding empirical Bayes estimator is $\etahat^{\rm ML} = \etahatsb(\muhat^{\rm ML}, \lambdaahat^{\rm ML}, \lambdabhat^{\rm ML})$ and will be referred to as EBMLE (for Empirical Bayes Maximum-Likelihood). 

\begin{lemma}
The ML estimates defined in \eqref{eq:ebmle-hyper} satisfy the following equations:
\begin{equation}
\begin{split}
\label{eq:mu-ml}
\mathrm{ \;\;I. } \qquad	& \muhat= \muhat_1 \cdot I\{\muhat_1 \in [\hat{a}_{\tau},\hat{b}_{\tau}] \} + \hat{a}_{\tau} \cdot I\{\muhat_1 < \hat{a}_{\tau}\} +  \hat{b}_{\tau} \cdot I\{\muhat_1 > \hat{b}_{\tau}\} \hspace{2cm}\\
&\text{ where, }  \muhat_1 = (\one^{\topp} \Vhatinv \y) / ( \one^{\topp} \Vhatinv \one)~.
\end{split}
\end{equation}
If $\muhat_1 \in [\hat{a}_{\tau},\hat{b}_{\tau}] $ and $\hat{\lambda}_a, \hat{\lambda}_b$ are both strictly positive, they satisfy
\bel{eq:lambda-ml}
\begin{aligned}
&\mathrm{ \;II. }	\qquad \tr( \Vhatinv \Za \Za^{\topp} ) -   {\sigma^{-2}} \, \y^{\topp} (I - \Phat)^{\topp} \Vhatinv \Za \Za^{\topp} \Vhatinv (I - \Phat) \y  = 0   \text{\hspace{2cm}} \label{eq:lambdaa-ml} \\
&\mathrm{III. }	\qquad \tr( \Vhatinv \Zb \Zb^{\topp} ) - {\sigma^{-2}} \, \y^{\topp} (I - \Phat)^{\topp} \Vhatinv \Zb \Zb^{\topp} \Vhatinv (I - \Phat) \y = 0~,
\end{aligned}
\eel 
where \quad $\Phat = \one ( \one^{\topp} \Vhatinv \one )^{-1} \one^{\topp} \Vhatinv$. 
\end{lemma}
\noindent The derivation is standard and provided in Section~\ref{append.estimating.eqn} of the supplements, which also contain the estimating equation for the case when $\muhat_1 \notin [\hat{a}_{\tau},\hat{b}_{\tau}]$. 
If the solution to the estimating equations \eqref{eq:lambda-ml} includes a negative component, adjustments are needed in order produce the maximum-likelihood estimates of the scale hyper-parameters \citep[see][Section 2.2b-iii for a discussion of the one-way case]{searle2001generalized}. 

\smallskip
\textbf{Estimation of hyper-parameters.} 
We propose an alternative method for estimating the shrinkage parameters. 
Following the approach of \citet{xie2012sure}, for fixed $\tau \in (0,1]$ we choose the shrinkage parameters by minimizing unbiased risk estimate (URE) over estimators $\etahatsb$ in $\mathcal{S}[\tau]$. 
By Lemma~\ref{lem.misc.2} of the supplements, an unbiased estimate of the risk of $\etahatsb$,
\bes
R_{r,c}(\etab, \etahatsb(\mu,\lambdaa,\lambdab)) \triangleq \frac{1}{rc} \ex\| \etahatsb(\mu,\lambdaa,\lambdab) - \etab \|^2 ,
\ees
is given by 
\bel{eq:sure}
\URE(\mu, \lambdaa, \lambdab) = \rcinv \big\{ \sigma^2 \tr(M) - 2\sigma^2 \tr (\Vinv M^2) + (\y - \one\mu)^{\topp} [\Vinv M^2 \Vinv] (\y - \one\mu) \big \}.
\eel
Hence we propose to estimate the tuning parameters of the class $\mathcal{S}[\tau]$ by
\bel{eq:sure-def}
(\muhat^{\rm U}, \lambdaahat^{\rm U}, \lambdabhat^{\rm U}) = \argmin_{\mu \in [\hat{a}_{\tau},\hat{b}_{\tau}], \lambdaa\geq 0, \lambdab\geq 0}  \URE( \mu, \lambdaa, \lambdab). 
\eel
The corresponding empirical Bayes estimator is $\etahatb^{\rm URE} = \etahatsb(\muhat^{\rm U}, \lambdaahat^{\rm U}, \lambdabhat^{\rm U})$. 
As in the case of maximum likelihood estimation, there is no closed-form solution to \eqref{eq:sure-def}, but we can characterize the solutions by the corresponding estimating equations.

\begin{lemma}
	The URE estimates of \eqref{eq:sure-def} statisfy the following estimating equations:
	\begin{equation}
	\begin{split}\label{eq:mu-sure}
	\mathrm{\;I.}\quad	& \muhat= \muhat_1 \cdot I\{\muhat_1 \in [\hat{a}_{\tau},\hat{b}_{\tau}] \} + \hat{a}_{\tau} \cdot I\{\muhat_1 < \hat{a}_{\tau}\} +  \hat{b}_{\tau} \cdot I\{\muhat_1 > \hat{b}_{\tau}\} \hspace{2.2cm}\\
		&\text{ where, } \muhat_1 = \big(\one^{\topp} [\Vhatinv M^2 \Vhatinv] \y\big) \big/ \big( \one^{\topp} [\Vhatinv M^2 \Vhatinv] \one\big)~. 
	\end{split}
	\end{equation}
	If $\muhat_1 \in [\hat{a}_{\tau},\hat{b}_{\tau}]$ and $\hat{\lambda}_a, \hat{\lambda}_b$ are both strictly positive, they satisfy:
	\bel{eq:lambda-sure}
	\begin{aligned}\label{eq:lambdaa-sure}
	&\mathrm{\;II.} \quad	\tr( \Vhatinv \Za \Za^{\topp} \Vhatinv M^2 ) - {\sigma^{-2}}\, \y^{\topp} (I - \Phat)^{\topp} \Vhatinv \Za \Za^{\topp} \Vhatinv M^2 \Vhatinv (I - \Phat) \y = 0 \text{\hspace{0.2cm}}\\
	&\mathrm{III.}	\quad \tr( \Vhatinv \Zb \Zb^{\topp} \Vhatinv M^2 ) - {\sigma^{-2}}\, \y^{\topp} (I - \Phat)^{\topp} \Vhatinv \Zb \Zb^{\topp} \Vhatinv M^2 \Vhatinv (I - \Phat) \y = 0~,
	\end{aligned}
	\eel
	where \quad $\Phat = \one ( \one^{\topp} [\Vhatinv M^2 \Vhatinv] \one )^{-1} \one^{\topp} \Vhatinv M^2 \Vhatinv$. 
\end{lemma}
\noindent The derivation is provided in Section~\ref{append.estimating.eqn} of the supplementary materials. 
Comparing the two systems of equations \eqref{eq:lambda-ml} and \eqref{eq:lambda-sure}  without substituting the value of $\mu$, it can be seen that the URE equation involves an extra term $\Vhatinv M^2$ in both summands of the left-hand side, as compared to the ML equation. 
The estimating equations therefore imply that the ML and URE solutions may differ when the design is unbalanced. 
In Section~\ref{sec:theory}, we show that the URE estimate $\etahatb^{\sf URE}$ is asymptotically optimal as $r,c \to \infty$, and the numerical simulations in Section~\ref{sec:simulation} demonstrate that in certain situations EBMLE performs significantly worse. 

\section{Estimation in Model with Missing Cells}\label{sec:missing}
A more general model than \eqref{eq:model-twoway} allows some cells to be empty. 
Hence, consider
\bel{eq:model-twoway-missing}
\begin{aligned}
y_{ij} &= \eta_{ij} + \epsilon_{ij} \quad \text{ for } \quad  (i,j) \in \mathcal{E}\\
\eta_{ij} &= \mu + \alpha_i + \beta_j \text{ and } \epsilon_{ij} \sim N(0,\sigma^2 K^{-1}_{ij})~,
\end{aligned}
\eel
where $\mathcal{E} = \{ (i,j): K_{ij} \geq 1 \} \subseteq \{ 1,...,r \} \otimes \{ 1,...,c \}$ is the set of indices corresponding to the nonempty cells. 
As before, $\sigma^2>0$ is assumed to be known. 
Our goal is in general to estimate all cell means that are {\it estimable} under \eqref{eq:model-twoway-missing} rather than only the means of observed cells. 
For ease of presentation and without loss of generality, from here on we assume that $\mathcal{E}$ is a connected design \citep{dey1986theory} so that all $r c$ cell means are estimable. 
%
\par
We will need some new notation to distinguish between $\ex[\y]\in \R^{|\mathcal{E}|}$ and the $rc$ vector consisting of all cell means. 
In general, the notation in \eqref{eq:model-bayes} is reserved for quantities associated with the observed variables. 
As before, $\thetab = (\alphab^{\topp},\betab^{\topp})^{\topp}$. 
The matrix $M = \text{diag}(K^{-1}_{ij}: (i,j) \in \mathcal{E})$, where the indices of diagonal elements are in lexicographical order. 
Let $\Ztil = [\one_{rc}\ \ I_R \otimes 1_C \ \ 1_R \otimes I_C]$  be the $rc \times (r+c+1)$ design matrix associated with the unobserved complete model. 
The $|\mathcal{E}|\times (r+c+1)$ ``observed" design matrix $Z$ 
is obtained from $\Ztil$ by deleting the subset of rows corresponding to $\mathcal{E}^c$. 
With the new definitions for $Z$ and $M$, we define $\V$ by \eqref{eq:V-compound}.
Finally, let $\etatil = \Ztil \thetab \in \R^{rc}$ be the vector of all estimable cell means and $\etab = Z \thetab \in \R^{|\mathcal{E}|}$ be the vector of cell means for only the observed cells of \eqref{eq:model-twoway-missing}. 
Hence, assuming $\mathcal{E}$ corresponds to connected design, we consider estimating $\etatil$  under the normalized sum-of-squares loss. 

Note that since $\etatil$ is estimable, it must be a linear function of $\etab$. 
The following lemma is an application of the basic theory of estimable functions and is proved in the Section~\ref{append.missing} of the supplementary materials.  
\begin{lemma}\label{lem.missing.1}
If $\etatil$ is estimable, then $\etatil = \Ztil (Z^{\topp}Z)^- Z^{\topp} \etab$, where $(Z^{\topp}Z)^-$ is any generalized inverse of $Z^{\topp}Z$.
\end{lemma}
\noindent In particular, writing $\Zinv$ for the Moore-Penrose pseudo-inverse of $Z$, we therefore have 
$
\etatil = \Ztil \Zinv \etab.
$
Thus, we can rewrite the loss function as
\bel{eq:loss.q1}
L_{r,c}(\etatil, \hat{\etatil}) \triangleq \frac{1}{rc} \| \hat{\etab}_{\sf c} - \etatil \|^2 = \frac{1}{rc} (\hat{\etab} - \etab)^{\topp} Q (\hat{\etab} - \etab) = L_{r,c}^Q(\etab, \hat{\etab})~,
\eel
where 
\bel{eq:qmat}
Q = (\Ztil \Zinv)^{\topp} \Ztil \Zinv. 
\eel
In other words, the problem of estimating $\etatil$ under sum-of-squares loss can be recast as the problem of estimating $\etab=\ex[\y]$ under appropriate quadratic loss. 
This allows us to build on the techniques developed in the previous section and extend their applicability to the loss in \eqref{eq:loss.q1}. 
The standard unbiased estimator of $\etatil$ is the weighted least squares estimator. 
The form of the Bayes estimator for $\etab$ under \eqref{eq:model-bayes} is not affected by the generalized quadratic loss $L^Q_{r,c}$ and is still given by \eqref{eq:bayes}, with $M, \Vinv$ as defined in the current section.
As before, for any pre-specified $\tau \in (0,1]$ we consider the class of estimators $\mathcal{S}[\tau]$  defined in \eqref{eq:bayes-family}. 
The EBMLE estimates the hyper-parameters $\mu,\lambdaa,\lambdab$ based on the marginal likelihood $\y$ according to \eqref{eq:model-bayes}, where $M, \Vinv$ are as defined in the current section. 
As shown in Lemma~\ref{lem.missing.3} of the supplements, an unbiased estimator of the point risk corresponding to \eqref{eq:loss.q1},
 \bes 
 R^Q_{r,c}(\etab, \etahatsb(\mu,\lambdaa,\lambdab)) \triangleq \ex \big \{ L^Q_{r,c}\big(\etab, \etahatsb(\mu,\lambdaa,\lambdab)\big)\big\}, 
 \ees
is given by
\begin{align}\label{eq:sure-missing}
\begin{split}
\UREQ (\mu, \lambdaa, \lambdab)  = (rc)^{-1} \big [\sigma^2 \tr(QM)  &- 2\sigma^2 \tr (\Vinv MQM) \\
&+ (\y - \mu \one)^{\topp} \big[\Vinv MQM \Vinv\big] (\y -  \mu \one) \big]. 
\end{split}
\end{align}
The URE estimates of the tuning parameters are 
\begin{align}\label{eq:sure-def-missing}
(\muhat^{\rm U_Q}, \lambdaahat^{\rm U_Q}, \lambdabhat^{\rm U_Q}) = \argmin_{\mu \in [\hat{a}_{\tau}, \hat{b}_{\tau}], \,\lambdaa\geq 0, \,\lambdab\geq 0}  \UREQ\big(\mu, \lambdaa, \lambdab\big)~,
\end{align}
and the corresponding EB estimate is $\etahatb^{\rm URE}=\etahatsb(\muhat^{\rm U_Q}, \lambdaahat^{\rm U_Q}, \lambdabhat^{\rm U_Q})$. 
Equivalently, the estimate for $\etatil$ is 
$ \etahatb_{\sf c}^{\rm URE} = \Ztil \Zinv \etahatsb(\muhat^{\rm U_Q}, \lambdaahat^{\rm U_Q}, \lambdabhat^{\rm U_Q})$. 
The estimating equations for the URE as well as ML estimates of $\mu, \lambdaa, \lambdab$ can be derived similarly to those in the all-cells-filled model. 

\section{Risk Properties and Asymptotic Optimality of the URE Estimator}\label{sec:theory}
We now present the results that establish the optimality properties of our proposed URE-based estimator.  
We  present the result for the quadratic loss $L_{r,c}^Q$ of the previous section with the matrix $Q$ defined in \eqref{eq:qmat}. Substituting $Q$ with $I_{rc}$ will give us the results for the fully-observed model \eqref{eq:model-twoway}, which are also explained. 
In proving our theoretical results we make the following assumptions: 

\bigskip

\noindent \textbf{A1. On the parameter space:} We assume that the parameter $\etab_c$ in the complete model is estimable and satisfies the following second order moment condition:
\begin{align}\label{assump.1}{(A1)}
	\qquad \lim_{r, c \to \infty} \frac{1}{rc} \sum_{i=1}^r \sum_{j=1}^c \eta_{i,j}^2 < \infty . 
\end{align}
This assumption is very mild, and similar versions are widely used in the EB literature (see Assumption $C'$ of \citealp{xie2012sure}). 
It mainly facilitates a shorter technical proof and can be avoided by considering separate analyses of the extreme cases. 
\\ 
\textbf{A2. On the design matrix:}
Denoting the largest eigenvalue of a matrix $A$ by $\lambda_1(A)$, the matrix $Q$  in \eqref{eq:qmat} is assumed to satisfy
\begin{align}\label{assump.2}{(A2)}
\qquad \lim_{r, c \to \infty} (rc)^{-1/8}\, ( \log (rc) )^2 \, \nu_{r,c} \, \lambda_1(Q) = 0~,
\end{align}  
$\text{ where } \nu_{r,c}=\max\{K_{ij}:(i,j)\in \mathcal{E}\}/\min\{K_{ij}:(i,j)\in \mathcal{E}\}$. 
As shown in Lemma~\ref{lem.assump.1} in the Appendix, $\lambda_1(Q)$ equals the largest eigenvalue of $(Z_c'Z_c)(Z'Z)^{\dagger}$. 
Intuitively, it represents the difference in information between the observed data matrix and the complete data matrix  $Z_c$. 
If there are many empty cells, $\lambda_1((Z_c'Z_c)(Z'Z)^{\dagger})$ will be large and may violate the above condition. 
On the contrary, in the case of the completely observed data we have $\lambda_1(Q)=1$ (see Lemma~\ref{lem.assump.1}). 
Thus, in that case the assumption reduces to $\lim_{r, c \to \infty} (rc)^{-1/8} (\log(rc))^2 \, \nu_{r,c} = 0$. 
This condition amounts to controlling in some sense the extent of imbalance in the number of observations procured per cell. Here, we are allowing the imbalance in the design to asymptotically grow to infinity but at a lower rate than $(rc)^{1/8}/(\log(rc))^2$. This assumption on the design matrix  is essential for our asymptotic optimality proofs. 
Section~\ref{append.theory} of the Appendix shows its role in our proofs and a detailed discussion about it is provided in the supplementary materials. 


\bigskip

\noindent \textbf{Asymptotic optimality results.} 
The following theorem forms the basis for the results presented in this section:
\begin{theorem} \label{thm:loss-approx-missing}
Under Assumptions A1-A2, with  $d_{r,c}= \thr^7\, \nu_{r,c}^3\,\lambda_1^3(Q)$ and $\thr = \log(rc)$ we have
	 $$ \lim_{\substack{\hspace{1mm}\\ r \to \infty \\[0.5ex] c \to \infty}} \;\; d_{r,c} \cdot \bigg\{ \sup_{\substack{|\mu| \leq \thr\\[0.5ex] \lambdaa, \lambdab\geq 0}} \ex \Big \vert \UREQ_{r,c}( \mu, \lambdaa, \lambdab) -  L_{r,c}^Q\big(\etab, \etahatsb(\mu, \lambdaa, \lambdab)\big) \Big\vert \bigg\} = 0.$$
\end{theorem}
Theorem \eqref{thm:loss-approx-missing} shows that the unbiased risk estimator approximates the true loss pointwise uniformly well over a set of hyper-parameters where $\lambdaa, \lambdab$ can take any non-negative value and the location hyper-parameter $\mu$ is restricted to the set $[-\thr,\thr]$,  which grows as $r, c$ increases.  
The set of all hyper-parameters considered in $\mathcal{S}[\tau]$ differs from the aforementioned set, as there $\mu$ was restricted to be in the data-dependent set $[\hat{a}_{\tau}, \hat{b}_{\tau}]$. However,  as $r , c \to \infty$. $[\hat{a}_{\tau}, \hat{b}_{\tau}]$  is asymptotically contained in $[-\thr,\thr]$ (see Lemma~\ref{lem:sec4.temp.1}), so  Theorem~\ref{thm:loss-approx-missing} asymptotically covers all hyper-parameters considered in $\mathcal{S}[\tau]$ for any $\tau \in (0,1]$.
This explains intuitively why in choosing the  hyper-parameters by minimizing an unbiased risk estimate as in \eqref{eq:sure-missing}, we can expect the resulting estimate $\etahatsb(\muhat^{\rm U_Q}, \lambdaahat^{\rm U_Q}, \lambdabhat^{\rm U_Q})$ to have competitive performance. 
To compare the asymptotic performance of our proposed estimate, we define the oracle loss (OL) hyper-parameter as:
\bes
\big( \ \mutil^{\rm OL}, \lambdaatil^{\rm OL}, \lambdabtil^{\rm OL} \ \big) = \argmin_{\mu \in [\hat{a}_{\tau},   \hat{b}_{\tau}]; \ \lambdaa, \ \lambdab\geq 0}  L^Q\big(\etab, \etahatsb(\mu, \lambdaa, \lambdab)\big)  
\ees
and the corresponding oracle rule 
\bel{eq:ol-missing}
\etatb^{\rm OL}_{\sf c} = \Ztil \Zinv \etahatsb( \mutil^{\rm OL}, \lambdaatil^{\rm OL}, \lambdabtil^{\rm OL} )~.
\eel
Note that the oracle rule 
depends on the unknown cell means $\etatil$ and is therefore not a ``legal" estimator. 
It serves as the theoretical benchmark for the minimum attainable error by any possible estimator: by its definition, no EB estimator in our class 
can have smaller risk than $\etatil^{\rm OL}$. 
The following two theorems show that our proposed $\text{URE}$-based estimator performs asymptotically {nearly as well as} the oracle loss estimator.  The results  hold for any class $\mathcal{S}[\tau]$ where $\tau \in (0,1]$. 
These results are in terms of the usual quadratic loss on the vector of all cell-means. 
Note that, based on our formulation of the problem in sections~\ref{sec:bayes} and \ref{sec:missing}, both theorems \ref{thm:asym-ol-pr-missing} and \ref{thm:asym-ol-ev-missing} simultaneously cover the missing and fully-observed model.  

\begin{theorem} \label{thm:asym-ol-pr-missing}
Under Assumptions A1-A2,  for any $\epsilon > 0$ we have
\bes
\displaystyle \lim_{  \substack{ \\r\to \infty \\ c\to \infty}  } P \big\{  L_{r, c} (\etatil, \etahatb_{\sf c}^{\rm URE}) \geq L_{r, c} (\etatil, \etatb_{\sf c}^{\rm OL}) + \epsilon \big\} = 0~. 
\ees
\end{theorem}
The next theorem asserts than under the same conditions, the URE-based estimator is asymptotically as good as the {oracle estimator} in terms of risk.  

\begin{theorem} \label{thm:asym-ol-ev-missing}
Under Assumptions A1-A2, the following holds:
\bes
\displaystyle \lim_{  \substack{ \\r\to \infty \\ c\to \infty}  }{  R_{r,c}(\etatil, \etahatb_{\sf c}^{\rm URE}) - \ex[ L_{r,c}(\etatil, \etatb_{\sf c}^{\rm OL}) ] } = 0 ~.  
\ees
\end{theorem}

Finally, as the oracle performs better than any empirical Bayes estimator associated with $\mathcal{S}[\tau]$, a consequence of the above two theorems is that that $\text{URE}$-based estimator cannot be improved by any other such empirical Bayes estimator. 

\begin{corollary} \label{cor:asym-opt-missing}
Under Assumptions A1-A2, it holds that for any estimator $\etahatsb(\muhat, \lambdaahat, \lambdabhat)$ corresponding to the class $\mathcal{S}[\tau]$  we have
\benu[(a)]
\item $\displaystyle \lim_{  { \\r\to \infty,\, c\to \infty}  } P \big\{  L_{r,c}^Q(\etab, \etahatb^{\rm URE}) \geq L_{r,c}^Q(\etab, \etahatsb(\muhat, \lambdaahat, \lambdabhat)) + \epsilon \big\} = 0 ~.$  

\item $\displaystyle \limsup_{  r \to \infty,\ c\to \infty  }   R^Q_{r,c}(\etab, \etahatb^{\rm URE}) - R^Q_{r,c}(\etab, \etahatsb(\muhat, \lambdaahat, \lambdabhat))  \leq 0~. $
\eenu
\end{corollary}
Unlike the above two theorems, this corollary is based on the quadratic loss $L^Q$. 
It emphasizes the nature of our estimating class $\mathcal{S}[\tau]$.  
In Section~\ref{sec:bayes} we saw that the EBMLE and URE generally produce different solutions in unbalanced designs; 
combined with Corollary \eqref{cor:asym-opt-missing}, this implies that, asymptotically EBMLE generally does not achieve the optimal risk of an EB estimator corresponding to the class $\mathcal{S}[\tau]$ (otherwise the EBML estimate for $\etab$ would have to be very close to the URE estimate). 

The proofs of theorems~\ref{thm:asym-ol-pr-missing} and \ref{thm:asym-ol-ev-missing} and that of Corollary~\ref{cor:asym-opt-missing} is left to Section~\ref{append.theory} of the Appendix. 
The proofs rely heavily on the asymptotic risk estimation result of Theorem~\ref{thm:loss-approx-missing},  
which in turn uses the asymptotic risk properties of estimators in $\mathcal{S}[\tau]$. Below, we sketch its proof by describing the interesting risk properties of these estimations. 
\par
To conclude this section, we would like to point out the qualitative differences between the type of results included in the current section and the results for the one-way normal mean estimation problem exemplified in \citet{xie2012sure} and especially point out the differences in the proof techniques. In estimation theory, the optimality of shrinkage estimators in one-way problems is usually studied through a sequence model (see Ch. 2 of \citealp{johnstone2011gaussian}), where there is a natural indexing on the dimensions in the parametric spaces. In unbalanced designs, the cell mean estimation problem in 2-way layouts cannot be reduced to estimating independent multiple vectors, and so there is no indexing on the parametric space under which the ``row" effects and the ``column" effects can be decoupled. Thus, the approach of \citet{xie2012sure}, which would require showing uniform convergence of the difference between the URE and the loss over the hyper-parametric space, i.e., showing $L_1$ convergence of    
 $ \sup_{\mu \in [\hat{a}_{\tau}, \hat{b}_{\tau}];\, \lambdaa, \lambdab\geq 0} \vert \UREQ_{r,c}( \mu, \lambdaa, \lambdab) -  L_{r,c}^Q(\etab, \etahatsb(\mu, \lambdaa, \lambdab))|$ to $0$, cannot be trivially adapted to the two-way layout. Instead, in Theorem~\ref{thm:loss-approx-missing} we show the pointwise convergence of the expected absolute difference between the URE and the loss. 
 Specifically, we show that as $r, c \to 0$, it converges at a rate $d_{r,c}$ uniformly over the essential support of the hyper-parameters. Using this pointwise convergence, its rate and the properties of the loss function (see Section~\ref{append:theory-2}), we prove the optimality results of
theorems \ref{thm:asym-ol-pr-missing}, \ref{thm:asym-ol-ev-missing},  which are of the same flavor as those in \citet{xie2012sure} for the one-way case. Our pointwise convergence approach greatly helps to tackle the difficulties encountered when passing to the two-way problem, where computations involving matrices are generally unavoidable. Our pointwise convergence result is proved by a moment-based concentration approach, which translates the problem into bounding moments of Gaussian quadratic forms involving matrices with possibly dependent rows and columns. 
The following two lemmas, which are used in proving Theorem~\ref{thm:loss-approx-missing}, display our moment-based convergence approach, where the concentration of relevant quantities about their respective mean is proved. To prove Theorem~\ref{thm:loss-approx-missing} we first show Lemma \ref{lem:risk-1}, in which the URE methodology estimates the risk in $L_2$ norm pointwise uniformly well for all estimators in $\mathcal{S}[\tau]$ that shrink towards the origin (i.e., with $\mu$ set at $0$). Thereafter, in Lemma \ref{lem:risk-2} we prove that  the loss of those estimators concentrate around their expected values (risk) when  we have large number of row and column effects. 
\begin{lemma}\label{lem:risk-1}
Under Assumptions~A1-A2, with  $d_{r,c}= \thr^7\, \nu_{r,c}^3\,\lambda_1^3(Q)$, $\thr= \log(rc)$,
	$$\lim_{\substack{\hspace{1mm}\\ r \to \infty \\[0.5ex] c \to \infty}} \;\; d_{r,c}^2 \cdot \bigg\{\sup_{\substack{ \lambdaa, \lambdab\geq 0}} \ex \Big [ \UREQ_{r, c}( 0, \lambdaa, \lambdab) - R^Q_{r,c}(\etab, \etahatsb(0, \lambdaa, \lambdab)) \Big]^2 \bigg\}= 0~.$$
\end{lemma}

\begin{lemma}\label{lem:risk-2}
Under Assumptions~A1-A2, with  $d_{r,c}= \thr^7\, \nu_{r,c}^3\,\lambda_1^3(Q)$, $\thr= \log(rc)$,
	$$\lim_{\substack{\hspace{1mm}\\ r \to \infty \\[0.5ex] c \to \infty}} \;\; d_{r,c}^2 \cdot \bigg\{\sup_{\lambdaa, \lambdab\geq 0} \ex \Big [ L_{r, c}^Q\big(\etab, \etahatsb(0, \lambdaa, \lambdab)\big) - R^Q_{r,c}(\etab, \etahatsb(0, \lambdaa, \lambdab)) \Big]^2 \bigg\} = 0~.$$ 
\end{lemma}

\noindent If we restrict ourselves to only estimators in $\mathcal{S}[\tau]$ that shrink towards the origin, then Theorem~\ref{thm:loss-approx-missing} follows directly from the above two lemmas. As such, for this subset of estimators, the lemmas prove a stronger version of the theorem with convergence in $L_2$ norm.  The proof is extended to general shrinkage estimators by controlling the $L_1$ deviation between the true loss and its URE-based approximation through the nontrivial 
use of the location invariance structure of the problem.  The proofs of all these results are provided in Section~\ref{append.theory} of the Appendix. The results for the weighted loss $L_{r,c}^{\text{wgt}}$ (defined in Section~\ref{sec:bayes}) are discussed in Section~\ref{append.weighted} of the supplements. 

%
%
\par


\section{URE in Balanced Designs} \label{sec:bal}
In this section we inspect the case of a balanced design, $K_{ij} = k,\ 1 \leq i \leq r,  1 \leq j \leq c$. 
We show that under a balanced design the problem essentially decouples into two independent one-way problems, in which case the URE and EBMLE estimates coincide \citep[see also][Section 2]{xie2012sure}. 
As a bonus, the analysis will suggest another class of shrinkage estimators for the general, unbalanced two-way problem by utilizing the one-way estimates of \citet{xie2012sure}. 

To carry out the analysis, suppose without loss of generality that $K=1$. 
Let the grand mean and the row and column main effects be
\bel{eq:main-effects-new} 
m = \mu + \alpha_{\cdot} + \beta_{\cdot},\ \ \ \ a_i = \alpha_i - \alpha_{\cdot},\ \ \ \ b_j = \beta_j - \beta_{\cdot} \quad \quad 
\eel
and let $\ba = (a_1,...,a_r)^{\topp}, \bb = (b_1,...,b_c)^{\topp}$. 
Then, in the balanced case, the Bayes estimator $\etahatsb(y_{\cdot \cdot}, \lambdaa, \lambdab)$, obtained by substituting the mean of $\y$ for $\mu$ in \eqref{eq:bayes}, is
\begin{align}
&\big \{ \etahatsb_{ij} (y_{\cdot \cdot}, \lambdaa, \lambdab) \big \} = \mhatls + c_{\alpha}(\lambdaa) \; \ahatls_i + c_{\beta}(\lambdab) \; \bhatls_j ~, \quad \label{eq:sure-bs} \qquad \qquad \qquad\\
&\text{ where} \quad \mhatls = y_{\cdot \cdot }, \ \ \ \ \ \ahatbls_i = y_{i\cdot} - y_{..} \;, \ \ \ \ \ \bhatbls_i = y_{\cdot j} - y_{..} \label{eq:ls-bal}
\end{align}
are the least squares estimators, and $c_{\alpha} := c_{\alpha}(\lambdaa)=\lambdaa/(\lambdaa+\sigma^2/c)$ and $c_{\beta} := c_{\beta}(\lambdab)=\lambdab/(\lambdab+\sigma^2/r)$ are functions involving, respectively, only $\lambdaa$ or only $\lambdab$. Its risk $R(\etab, \etahatsb(y_{\cdot \cdot}, \lambdaa, \lambdab))$ decomposes as
\begin{align}
 \ex\Big\{  (\mhatls - m)^2  \Big\} + \frac{1}{r} \ex\Big\{  \sum_{i=1}^r(c_{\alpha} \ahatls_i - a_i)^2 \Big\} + \frac{1}{c} \ex\Big\{  \sum_{j=1}^c(c_{\beta} \bhatls_j - b_j)^2 \Big\} \label{eq:lin}
\end{align}
due to the orthogonality of the vectors corresponding to the three sums-of-squares (detailed derivation is provided in the supplements). 
Consequently, one obtains URE by writing URE for each of summands above. 
Moreover, since 
\bel{eq:ls-marginal}
\mhatls \sim N(m, \sigma^2 \lambda^2_m),\ \ \ \ \ \ahatbls \sim N_r(a, \sigma^2 \Lambda_a),\ \ \ \ \ \bhatbls \sim N_c(b, \sigma^2 \Lambda_b), 
\eel
minimizing URE jointly over $(c_{\alpha}, c_{\beta})$ therefore consists of minimizing separately the ``row" term over $c_{\alpha}$ and the ``column" term over $c_{\beta}$. 
Each of these is a ``one-way" Gaussian homoscedastic problem, except that the covariance matrices $\Lambda_{\alpha}, \Lambda_{\beta}$ are singular because the main effects are centered. The unbiased risk estimator will naturally take this into account and will possess the ``correct" degrees-of-freedom. 

The maximum-likelihood estimates for the two-way random-effects additive model do not have a closed-form solution even for balanced data \citep[][Ch. 4.7 d.]{searle2009variance}, so it is not possible that they always produce the same estimates as discussed above. 
On the other hand, the REML estimates coincide with the positive-part Moments method estimates \citep[][Ch. 4.8]{searle2009variance}, which, in turn, reduce (for known $\sigma^2$) to solving separately two one-way problems involving $\ahatbls$ for the rows and $\bhatbls$ for the columns. 
These have closed-form solutions and are easily seen to coincide with the URE solutions. 

In the unbalanced case, \eqref{eq:sure-bs} no longer holds, and so the Bayes estimates for $\ba$ and $\bb$ are each functions of both $\ahatbls$ and $\bhatbls$. We can nevertheless use shrinkage estimators of the form \eqref{eq:sure-bs} and look for ``optimal" constants $c_{\alpha} = c_{\alpha}(\lambdaa)$ and $c_{\beta} = c_{\beta}(\lambdab)$. 
Appealing to the asymptotically optimal one-way methods of \citet{xie2012sure}, we  consider the estimator 
\begin{align}
&\etahat^{\rm{XKB}}_{ij} = \mhatls + \chat^{\rm{XKB}}_{\alpha} \;\ahatls_i + \chat^{\rm{XKB}}_{\beta} \; \bhatls_j,\ \ \ 1\leq i \leq r,\ 1\leq j \leq c ~, \qquad  \qquad \\ \label{eq:xkb-twoway}
& \text{ where, } \; \chat^{\rm{XKB}}_{\alpha} = \argmin_{c_{\alpha} \in [0,1]} \URE \Big\{  \sum_{i=1}^r(c_{\alpha} \ahatls_i - a_i)^2 \Big\}, \\
& \qquad \qquad \chat^{\rm{XKB}}_{\beta} = \argmin_{c_{\beta}  \in [0,1]} \URE \Big\{  \sum_{j=1}^c(c_{\beta} \bhatls_j - b_j)^2 \Big\}~. \label{eq:xkb-twoway-where}
\end{align}
A slight modification of the parametric SURE estimate of \citet{xie2012sure} that shrinks towards 0 is required to accommodate the covariance structure of the centered random vectors $\ahatbls, \bhatbls$. 
Contrasting the performance of the optimal empirical Bayes estimators corresponding to this class of shrinkage estimators with that corresponding to the class $\mathcal{S}[\tau]$ of EB estimators
can be taken to quantify the relative efficiency of using one-way methods in the two-way problem.

\section{Computation of the URE Estimator} \label{sec:computation}
To compute the hyper-parameter estimates by the URE method, one could attempt to solve the estimating equations in \eqref{eq:lambda-sure}, which have no closed-form solution.  For example, one could fix the value of $\lambdaa$ to some initial positive value and solve the first equation in $\lambdab$.
Then, plug the solution into the second equation and solve for $\lambdaa$, and keep iterating between the two equations until convergence. 
If this approach is taken, a non-trivial issue to overcome will be obtaining the actual minimizing values $\lambdaa$ and $\lambdab$ when one of the solutions to \eqref{eq:lambda-sure} is negative. 
Another issue will be ascertaining the global optimality of the solutions, as $\URE$ is not necessarily convex in $(\mu, \lambdaa, \lambdab)$. To bypass these issues, we minimize $\URE$ by conducting a grid search on the scale hyper-parameters, and $\mu$ is subsequently estimated by \eqref{eq:mu-sure}. 
\par
A major hindrance for computations in large designs is the occurrence of the $(rc)\times (rc)$ matrix $\V^{-1}$, which depends on $\lambdaa$ and $\lambdab$. Inverting it can be a prohibitive task for even moderately large values of $r$ anc $c$, and it would need inversion at every point along the grid for a naive implementation. 
In our implementation, we adopt some of the key computational elements from the \texttt{lme4} package [Sec. 5.4 \citealp{bates2010lme4}] and produce an algorithm that works as fast as the computation of the EBMLE estimate with the \texttt{lme4} R-package. For the case of no empty cells, the pivotal step in our implementation is the representation of the $\URE$ criterion by the following expression:
\begin{align}\label{eq.temp6}
\URE = (rc)^{-1}\big[-\sigma^2 \tr(M) + 2 \sigma^2 \tr\{ (\Lambda^{\topp} Z^{\topp} M^{-1} Z \Lambda + I_q )^{-1} (\Lambda^{\topp} Z^{\topp} Z \Lambda) \} \\
 + \| M V^{-1} (\y - \one \mu) \|^2\big]~. 
\end{align}
The detailed steps for deriving \eqref{eq.temp6} are provided in Section~\ref{append.computation} of the appendix where the compuation of each of the above terms is also elaborately explained. 
 \eqref{eq.temp6} is numerically minimized jointly over $(\lambdaa,\lambdab)$, where the key step in evaluating it for a particular pair $(\lambdaa,\lambdab)$ is employing a sparse Cholesky decomposition for the matrix $\Lambda^{\topp} Z^{\topp} M^{-1} Z \Lambda + I_q$. 
This decomposition takes advantage of the high sparsity of $\Lambda^{\topp} Z^{\topp} M^{-1} Z \Lambda + I_q$.
It first determines the locations of non-zero elements in the Cholesky factor, which do not depend on the values of $(\lambdaa,\lambdab)$ and hence this stage is needed only once during the numerical optimization. 
This is the only costly stage of the decomposition and determining the values of the non-zero components is repeated during the numerical optimization. 
For the empty-cells case, the implementation is very similar after using the reduction to quadratic loss in $L^Q$ described in Section~\ref{sec:missing}.

\section{Simulation Study} \label{sec:simulation}
We carry out numerical experiments to compare the performance of the URE based estimator to that of different cell means estimators discussed in the previous sections. 
As the standard technique we consider the weighted Least Squares estimator $\etahatb^{\rm LS} = \muhat^{\rm LS} \one + Z\widehat{\thetab}^{\rm LS}$, where $(\muhat^{\rm LS}, \widehat{{\thetab}}^{\rm LS})$ is any pair that minimizes
\bes
(\y - \mu \cdot \one - Z\thetab)^{\topp} M^{-1} (\y - \mu \cdot \one - Z \thetab). 
\ees
The two-way shrinkage estimators reported are the maximum-likelihood empirical Bayes (EBML) estimator $\etahatb^{\rm ML}$ and the URE based estimator $\etahatb^{\rm URE}$, as well as versions of these two estimators which shrink towards the origin (i.e., with $\mu$ fixed at 0); these are designated in Table \ref{tab:relative-risk} as ``EBMLE (origin)" and ``URE (origin)". 
We also consider the generalized version of $\etahatb^{\rm XKB}$ discussed in Section~\ref{sec:bal}  which shrinks towards a general data-driven location and estimates the scale hyper-parameters based on two independent one-way shrinkage problems. For a benchmark we consider the oracle rule $\etahatb^{\rm OL}=\etahatsb(\mutil, \lambdaatil, \lambdabtil)$ where,
\bel{eq:ol-risk}
\big( \ \mutil, \lambdaatil, \lambdabtil \ \big) = \argmin_{\mu, \lambdaa\geq 0, \lambdab\geq 0} \big\| \y - M \Vinv (\y - \mu \cdot \one) -\etab \big\|^2~.
\eel
Since for any $\y$ the oracle rule minimizes the loss over all members of the parametric family \eqref{eq:bayes-family}, its expected loss lower bounds the risk achievable by any empirical Bayes estimator of the form \eqref{eq:bayes}. 

\smallskip
\textbf{Simulation setup.} 
We report results across 6 simulation scenarios. For each of them,  we draw $(\alphab, \betab, M^{-1}=\text{diag}(K_{11}, K_{12}, ..., K_{rc}))$ jointly from some distribution such that the cell counts $K_{ij}$ are i.i.d. and $(\alphab, \betab)$ are drawn from some conditional distribution given the $K_{ij}$s. 
We then draw $y_{ij} \sim N(\mu + \alpha_i + \beta_j, \sigma^2 K_{ij})$ independently, fixing $\mu = 0$ throughout and setting $\sigma^2$ to some (known) constant value. 
This process is repeated for $N=100$ time for each pair $(r,c)$ in a range of values, and the average squared loss over the $N$ rounds is computed for each of the estimators mentioned above. With 100 repetitions,  the standard error of the average loss for each estimator is at least one order-of-magnitude smaller than the estimated differences between the risks; hence, the differences can be safely considered significant.
The URE estimate is computed using the implementation described in Section~\ref{sec:computation}, and the oracle ``estimate" is computed employing a similar technique. 
The EBMLE estimate is computed using the \texttt{R} package \texttt{lme4} \citep{lme4}. 

\begin{figure}[H]
	\centering
	\includegraphics[width=\textwidth,height=0.85\textheight]{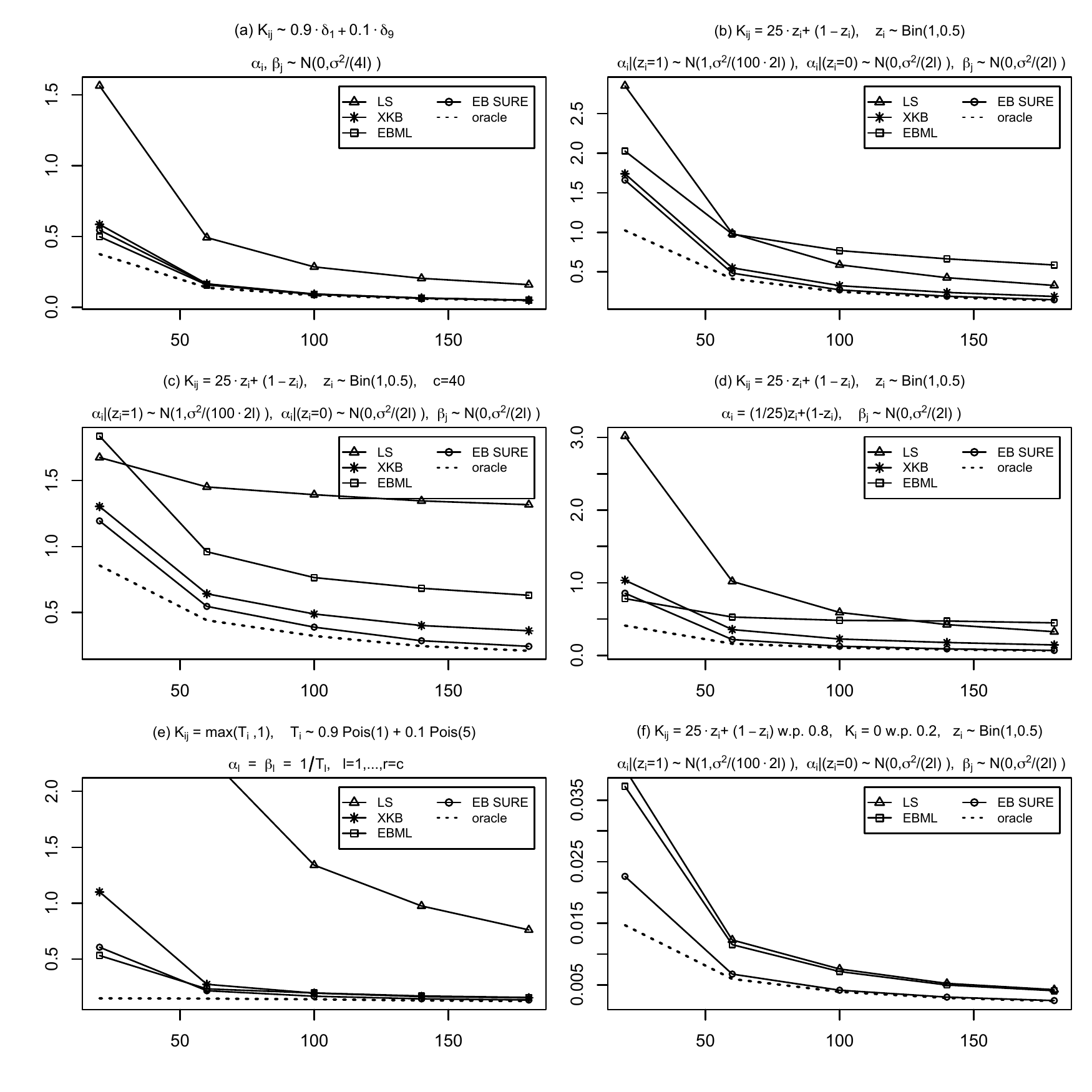}
	\caption{Risk of the various estimators in the six simulation scenarios of Table~\ref{tab:relative-risk}. The ordinate shows the risk of the estimators while we vary $L$ along the abscissa. Recall, $L=r=c$
		for all experiments in the table except (d) where $L=r$ and $c$ was fixed at $40$. .
	}
	\label{fig:simulation}
\end{figure}

\smallskip
Table~\ref{tab:relative-risk} shows the estimation errors of different estimators as a fraction of the estimated risk of the Least Squares (LS) estimator. 
We have equal number of row and column levels for all experiments except for scenario (c). 
In Figure~\ref{fig:simulation}, we have the plot of the mean square errors (MSE) of the URE, EBMLE, LS and the Oracle loss (OL) estimators across the six experiments as the number of levels in the design varies. 
It shows how the estimation errors of the different estimators compare with the minimum achievable (oracle) error rates as the number of levels in the designs increases. 
The general pattern reflected in the subplots shows an initial sharp decline with a gradual flattening-out of the error rates as the number of levels exceeds $100$, suggesting a setting within the asymptotic regime. 
In all the examples, the performance of our proposed URE based method is close to that of the oracle when the number of levels is large; 
for levels greater than $60$, there is no other estimator which is much better at any instance than the URE. 
On the contrary, in all examples except scenario (a) the EBMLE performs quite bad, and gets outperformed even by the ``one-way" XKB estimator. 
In cases with dependency between the effects and the cell counts, even the LS estimator can be preferable to the EBMLE (experiments (b) and (d)).  

\begin{table}[h]
\centering
\begin{tabular}{lcccccc}
  \hline
 & (a) & (b) & (c) & (d) & (e) & (f) \\ 
  \hline
  \text{LS} & 1.00 & 1.00 & 1.00 & 1.00 & 1.00 & 1.00 \\ 
   \text{EBMLE} & 0.31 & 1.79 & 0.48 & 1.37 & 0.21 & 0.96 \\ 
   \text{URE} & {\bf 0.31} & {\bf 0.45} & {\bf 0.19} & {\bf 0.21} & {\bf 0.18} & {\bf 0.58} \\ 
   \text{EBMLE (origin)} & 0.31 & 0.69 & 0.45 & 1.42 & 0.58 & 0.95 \\ 
   \text{URE (origin)} & 0.31 & 0.46 & 0.20 & 0.53 & 0.57 & 0.63 \\ 
   \text{XKB} & 0.31 & 0.58 & 0.28 & 0.44 & 0.20 & -  \\ 
   \text{Oracle} & 0.30 & 0.42 & 0.16 & 0.20 & 0.17 & 0.56 \\ 
   \hline
\end{tabular}
\caption{Estimation errors relative to the Least Squares (LS) estimator. The columns in the table correspond to the six simulation examples described in section~\ref{sec:simulation}.}\label{tab:relative-risk}
\end{table}

\noindent \textit{(a) Hierarchical Gaussian Model.} For $L\in \{ 20, 60,...,180 \}$ we set $r=c = L$ and $\sigma^2=25$. 
$K_{ij}$ are independent such that $P(K_{ij} = 1) = 0.9$ and $P(K_{ij} = 9) = 0.1$. 
For $1\leq i,j \leq L$, $\alpha_i, \beta_j$ are drawn from a $N(0, \sigma^2/(4L))$ distribution independently of the $K_{ij}$s. 
The joint distribution of the row effects, column effects and the $K_{ij}$s in this example obeys the Bayesian model under which the parametric estimator \eqref{eq:bayes} is derived. 
Hence the true Bayes rule is of that form, and the EBMLE is expected to perform well estimating the hyperparameters from the marginal distribution of $\y$. 
Indeed, the risk curve of the EBMLE approaches that of the oracle rule and seems to perform best for relatively small value of $L$. 
The MSE of the URE estimator, however converges to the oracle risk as $L$ increases. 
Interestingly, the performance of the XKB estimator seems to be comparable to that of URE and EBMLE for large values of $L$.

\noindent \textit{(b) Gaussian model with dependency between effects and cell counts.} For $L\in \{ 20, 60,...,180 \}$ we set $r=c = L$ and $\sigma^2=25$. 
In this example the $K_{ij}$ are no longer independent of the random effects. 
We take $K_{ij} = 1\cdot (1-Z_i) + 25 \cdot Z_i$ where $Z_i \sim Bin(1,0.5)$ independently, so that the cell frequencies are constant in each row. 
If $Z_i = 1$, $\alpha_i$ is drawn from a $N(1,\sigma^2/(100\cdot 2L))$ distribution, and otherwise from a $N(0, \sigma^2/(2\cdot L))$ distribution. 
$\beta_j$ are drawn independently from a $N(0, \sigma^2/(2 L))$ distribution. 
The advantage of our URE method over the EBMLE is clear in Figure~\ref{fig:simulation}; 
in fact, even the LS estimator seems to do better than the EBMLE for the values of $L$ considered here, a consequence of the strong dependency between the cell frequencies and the random effects. 
Again the XKB estimator performs surprisingly well. 

\noindent \textit{(c) Scenario (b) for different number of row and column effects.} This example is the same as example (b), except that we fix $c=40$ throughout and study the performance of the different estimators as number of row levels $r=L\in \{ 20, 60,...,180 \}$ varies. 
The performance of the LS estimator relative to the other methods is much worse than in the previous examples. 
The performance the URE estimator gets closer to that of the oracle as $r=L$ increases. 
The MSE of the XKB is significantly higher than that of the URE but much lower than that of the EBMLE.

\noindent \textit{(d) Non-Gaussian row effects.}  
For $L\in \{ 20, 60,...,180 \}$ we set $r=c = L$ and $\sigma^2=25$. 
In this example the row effects are {\it determined} by the $K_{ij}$. 
We take $K_{ij} = 1\cdot (1-Z_i) + 25 \cdot Z_i$ where $Z_i \sim Bin(1,0.5)$ independently, and set $\alpha_i = 1\cdot (1-Z_i) + (1/25) \cdot Z_i$. 
$\beta_j$ are drawn independently from a $N(0, \sigma^2/(2L))$ distribution. 
The URE estimator performs significantly better than the other estimators for large values of $L$, with about $50\%$ smaller estimated risk for $L=180$ than that of the XKB estimator, and even  much better compared to the other methods. 

\noindent \textit{(e) Correlated Main Effects.} For $L\in \{ 20, 60,...,180 \}$ we set $r=c = L$ and $\sigma^2=25$. 
In this example both the row and the column effects are determined by the $K_{ij}$. 
The cell frequencies $K_{lj} = \max(T_l, 1),\ 1\leq l \leq L, 1\leq j \leq L$, where $T_l, \ 1\leq l \leq L$,  are drawn independently from a mixture of a $Poisson(1)$ and $Poisson(5)$ distributions with weights $0.9$ and $0.1$, respectively. 
The row and column effects are $\alpha_l, \beta_l = 1/T_l,\ 1\leq l \leq L$. 
The MSE of the URE estimator is smaller than that of EBMLE by $14.7\%$ ($\widehat{\text{sd}}\text{(diff)} < 4\cdot 10^{-5}$) for $L=200$, but difference is not as big as in previous  examples. 
The LS estimator performs considerably worse than the rest.
\noindent \textit{(f) Missing Cells.} In the last example we study the performance of the estimators when some cells are empty. 
The setting is exactly as in example (b), except that after the $K_{ij}$ are drawn, each $K_{ij}$ is independently set to 0 (corresponding to an empty cell) with probability $0.2$. 
In accordance with the theory, the performance of the URE estimator approaches the oracle loss, and for $L=180$ achieves significantly smaller risk than that of the EBMLE, although not as significantly smaller as in example (b) with all cells filled ( $40\%$ vs $75\%$ smaller than EBMLE for examples (f) and (b), respectively). 
The performance of the LS estimator is comparable to that of the EBMLE. 
The XKB estimator is not considered here as it is not applicable when some data are missing.

\section{Real Data Example}\label{sec:real.data}
We analyze data collected on Nitrate levels measured in water sources across the US. 
Nitrates are chemical units found in drinking water that may lead to adverse health effects. 
According to the U.S. Geological Survey (USGS), excessive nitrate levels can result in restriction of oxygen transport in the bloodstream. 
The data was obtained from the Water Quality Portal cooperative (\url{http://waterqualitydata.us/}). 

\begin{figure}
  \centering
      \includegraphics[width=.45\textwidth]{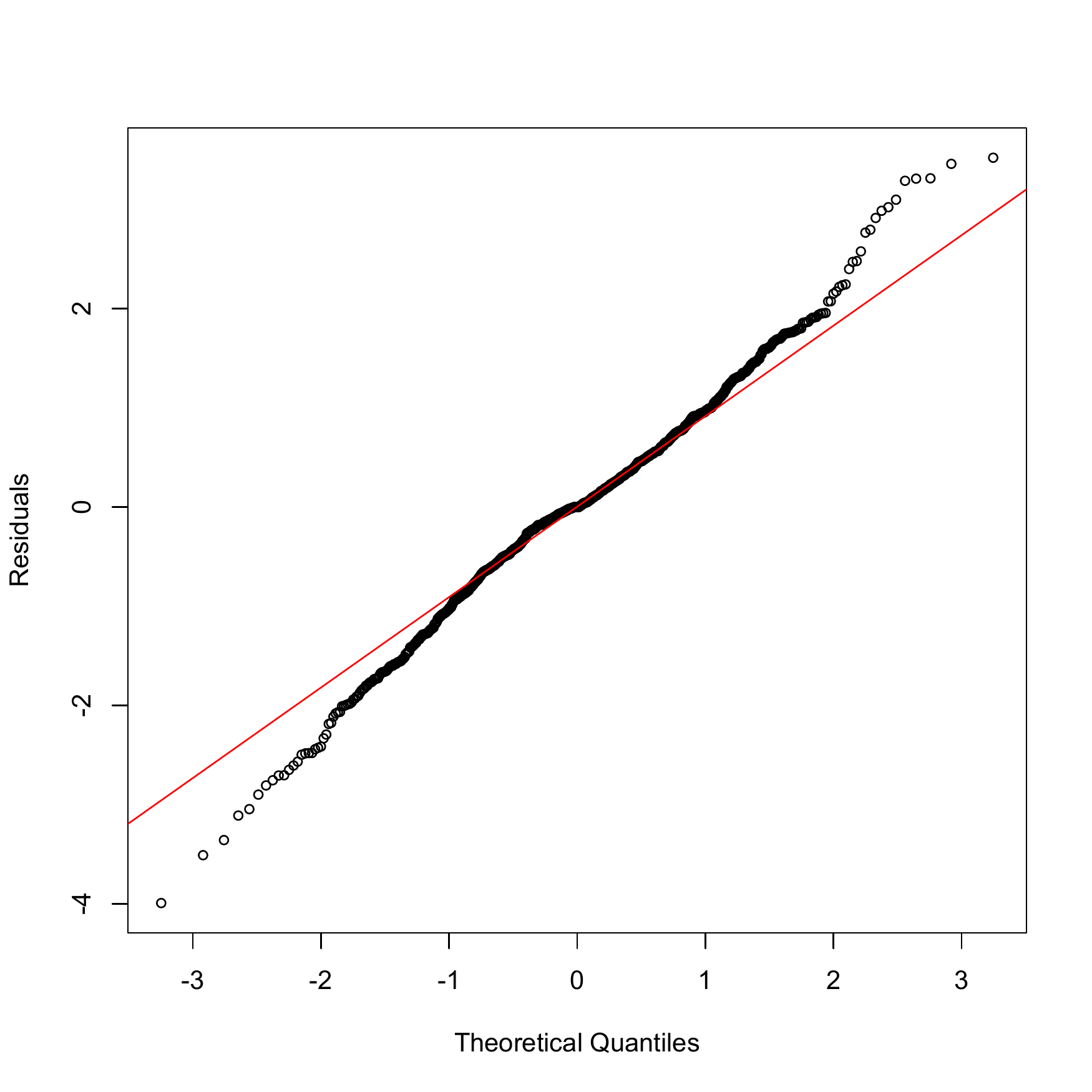}
      \includegraphics[width=.45\textwidth]{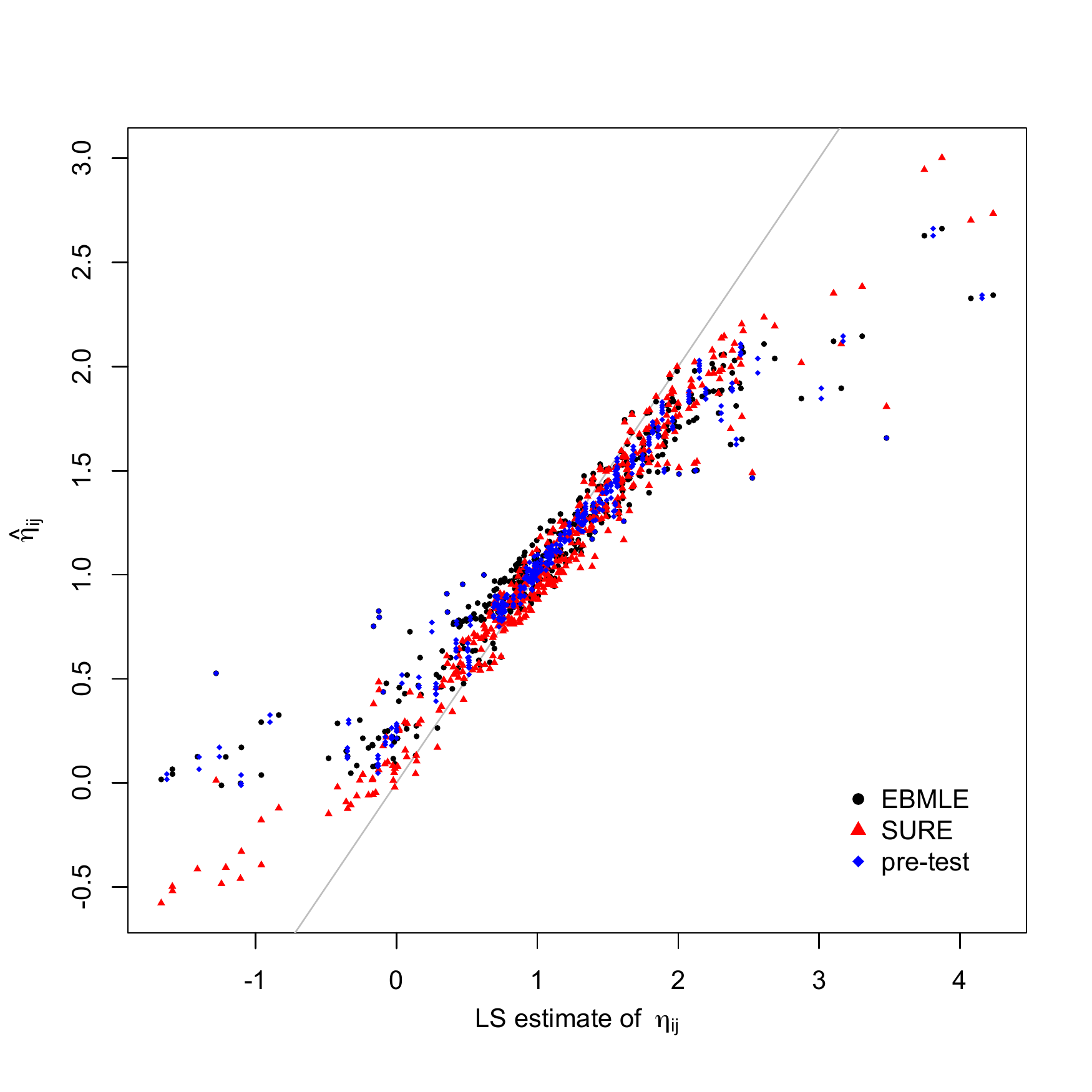}
  \caption[]{
\textit{Left:} Normal Q-Q plot for the residuals of the LS fit to the two-way model for water data. \textit{Right:} Plot of Shrinkage estimates vs. LS estimates of the cell means. 
The horizontal coordinate is the LS estimate and the vertical coordinate is an alternative estimate: EBMLE, URE or LS-county. 
EBMLE exhibits most shrinkage. 
The gray line is the identity line.
  }
  \label{fig:qq}
\end{figure}

We consider estimating the average Nitrate levels based on the location of the water resource and time when the measurement was taken. 
Specifically, we fit the homoscedastic Gaussian, additive two-way model
\begin{equation} \label{eq:water-model}
y_{ijk} = \eta_{ij} + \epsilon_{ijk}, \ \ \ \ \eta_{ij} = \mu + \alpha_i + \beta_j \ \ \ \ \ \ \ \ \ \ k=1,...,K_{ij}
\end{equation}
where $\alpha_{i}$ is the effect associated with the $i$-th level of a categorical variable indicating the hour of the day when the measurement was taken (by rounding to the past hour, e.g., for 14:47 the hour is 14); $\beta_j$ is the effect associated with the $j$-th US county; and $y_{ijk}$ is the corresponding log-transformed measurement of Nitrate level (in mg/l). 
The errors $\epsilon_{ijk}$ are treated as i.i.d. Gaussian with a fixed (known) variance equal to the the LS estimate $\hat{\sigma}^2$.  
We used records from January and February of 2014, and concentrated on measurements made between 8:00 and 17:00 as those were the most active hours. 
This yielded a total of 858 observations categorized into 9 different hour-slots (8-16) and 108 counties across the entire country. 
The data is highly unbalanced: 57\% of the cells are empty, and the cell counts among the nonempty cells vary between 1 to 12. 
Figure \ref{fig:qq} (left panel) shows the residuals from the standard LS fit for the data (note that this assumes independence of the noise terms). 
The alignment with the normal quantiles is better around the center of the distribution. 

A two-way Analysis-of-Variance yielded a highly significant p-value for county ($<10^{-5}$) but not for hour ($0.25$), for comparing the models with an without each variable (i.e.,  using Type II sums of squares). 
For the estimation problem, we considered the two-way shrinkage estimators, EBMLE and URE, as well as the ``pre-test" estimator which, failing to reject the null hypothesis for the overall effect of hour, proceeds with fitting the one-way LS estimate by county. 
We will refer to the latter as the ``one-way" estimator or as ``LS-county". 
As a two-way estimator, it can be interpreted as shrinking all the way to zero on hour, while providing no shrinkage at all for county. 
The ``usual" estimator is the LS estimator based on \eqref{eq:water-model}. 

Applying the shrinkage estimators to the entire data set, we observe that both shrink the LS estimates, but the shrinkage factors are quite different. 
Table \ref{tab:hyper} shows the estimates of the relative variance components $\lambdaa$ and $\lambdab$, corresponding to hour and county, respectively, as well as the estimates of the fixed term $\mu$, for each of the shrinkage estimators. 
There is a marked difference between the two methods in the estimates of the two variance components. 
Figure \ref{fig:qq} displays fitted values based on the two competing methods, as well as the one-way estimator (LS-county), against the corresponding LS estimate. 
In terms of shrinkage magnitude, it seems that EBMLE exhibits the most shrinkage among the three, and URE the least among the three, although the differences are not very big. 
Note that the individual shrinkage patterns could not be immediately anticipated from the values in Table \ref{tab:hyper} because of the imbalance in the data. 

\begin{table}[ht]
\centering
\begin{tabular}{lccc}
  \hline
 & $\mu$ & county & hour \\ 
  \hline
 \text{EBMLE} & 1.10 & 0.57 & 0.05 \\ 
 \text{URE} & 0.78 & 0.07 & 0.80 \\ 
   \hline
\end{tabular}
  \caption[]{
Estimated fixed effect ($\mu$) and relative shrinkage factors. 
  }   
  \label{tab:hyper}
\end{table}

 To compare the performance of the different estimators we carried out two separate analyses. 
In the first one, we split the data evenly and used the first portion for estimation and the second portion for validation. 
The second analysis is a data-informed simulation intended to compare performance of the estimators when the additive model \eqref{eq:water-model} is correctly specified. 

We begin with comparing the predictive performance against a holdout set. 
Recall that in the case of missing cells our aim is to estimate the vector $\etatil$ of {\it all estimable} cell means. 
For a random even split of the data into two subsets $\y^{(1)}, \y^{(2)}$, denote by $\hetatil^{(1)}$ an estimate of $\etatil$ based on $\y^{(1)}$ and denote by $\hetatil^{\text{\tiny LS}(2)}$ the {\it Least Squares} estimate of $\etatil$ based on $\y^{(2)}$.
As reflected in notation, we assume that the set of estimable cells is the same for the two portions. 
Then under \eqref{eq:water-model}, $\hetatil^{\text{\tiny LS}(2)}$ is an unbiased estimator of $\etatil$ and 
\begin{equation} \label{eq:sspe}
\rm{SSPE}[\hetatil^{(1)}] = \| \hetatil^{(1)} - \hetatil^{\text{\tiny LS}(2)}\|^2
\end{equation}
is the Sum of Squared Prediction Error of $\hetatil^{(1)}$. 
Instead of averaging \eqref{eq:sspe} directly over random splits, we could use the average of the estimated Total Squared Error
\begin{equation*}
\widehat{\rm{TSE}}[\hetatil^{(1)}] = \rm{SSPE}[\hetatil^{(1)}] - R(\etatil, \hetatil^{\text{\tiny LS}(2)})
\end{equation*}
where for any fixed split $R(\etatil, \hetatil^{\text{\tiny LS}(2)}) = \text{tr}[\text{Cov}(\Ztil \Zinv\hetatil^{\text{\tiny LS}(2)})]$ and is as an unbiased estimator of the expected risk of $\hetatil^{(2)}$ under a random even split (assuming that $\hat{\sigma}^2$ is the true variance). 
Unlike in the other sections we use the un-normalized sum-of-squares loss here, but this will not make any difference because {\it relative} estimated risks are compared. 
Note that under \eqref{eq:water-model} the average of $\| \hetatil^{\text{\tiny LS}(1)} - \hetatil^{\text{\tiny LS}(2)}\|^2 / 2$ over random splits of the data  is an unbiased estimator of the expected risk of $\hetatil^{\text{\tiny LS}(2)}$ under a random even split; we use it for our calculations in place of $R(\etatil, \hetatil^{\text{\tiny LS}(2)})$ to allow more flexibility in case of departures from the assumed model.

%



The first row of Table \ref{tab:risk} shows the average $\widehat{\rm{TSE}}$ for the two shrinkage estimators and the one-way estimator, as fraction of $\widehat{\rm{TSE}}_{\text{LS}}$, the average $\widehat{\rm{TSE}}$ for the LS estimator $\hetatil^{\text{\tiny LS}(1)}$, over $N=1000$ random splits of the data. 
We removed from the analysis all counties for which there was a total of less than 8 observations, and recorded the estimated TSEs for each of the $N$ rounds where the 
random split resulted in the same set of estimable cells for the two portions of the split. 
Hence the averages (and standard errors) are based on a slightly smaller effective number of simulation rounds, $N'=927$. 

Both shrinkage estimators show significant improvement over LS in terms of estimating the cell means. 
The EBMLE performs slightly better, with TSE 16\% smaller than URE. 
The estimated relative risk of the one-way estimator is smaller than LS but bigger than the two (empirical) linear shrinkage methods. 
The pre-test estimator is known to be dominated by a positive-part James-Stein estimator, and, for small values of the parameter, to perform better than the standard (LS) estimator \citep{sclove1972non}; 
this assumes balanced design, a correctly-specified model, and would entail testing the `preliminary' hypothesis at each round to decide whether to use the one- or two-way LS; none of these is exactly true of the current analysis, but the outcome of our analysis (also of the simulation analysis, reported next, in which at least misspecification is not a concern) is still in some informal sense consistent with the theoretical results.


\begin{table}[ht]
\centering
\begin{tabular}{l c c c}
  \hline
 & EBMLE & URE & LS-county\\ 
  \hline
validation & {\bf 0.42} & 0.5 & 0.72 \\ 
   \hline
simulation & 0.81 & {\bf 0.72} & 0.98 \\
  \hline
\end{tabular}
\caption[]{Estimated relative ${\rm{TSE}}$ for various estimators.
The first row of the table corresponds to analysis with validation. 
The second row corresponds to the data-informed simulation, in which data was simulated according to the additive model. 
Standard errors are $< 0.005$. 
The URE method seems to perform better under the assumed additive model. 
} 
\label{tab:risk}
\end{table}

As the estimators discussed in this paper are designed for the additive model \eqref{eq:water-model}, for our second analysis we compare the performance of the different methods (LS, LS-county, EBMLE and URE) when the data is actually generated from the additive model. 	
We set the LS estimate $\bfeta^{\text{\tiny LS}}$ for the model \eqref{eq:water-model} and the corresponding $\sigmahat^2$-- based on all 858 observations from all 108 counties -- as the ``truth", then draw an independent vector 
$\y^* \sim N_n(\bfeta^{\text{\tiny LS}}, \hat{\sigma}^2 I),\, n=\sum_{i,j} K_{ij},$
and compute the sum of squared loss $\| \hetatil^* - \etatil \|^2$ for each estimator $\hetatil^*$, where the asterisk indicates that the estimate is based on $\y^*$ only.  
This process was repeated $N=500$ times. 
The second row of Table \ref{tab:risk} shows the estimated risk of the two shrinkage estimators and the one-way estimator as a fraction of the risk of LS. 
All three estimators have higher risks (relative to LS) compared to the previous analysis, and the URE now has estimated relative risk about 10\% smaller than EBMLE. 
The one-way estimator now barely improves over the standard LS estimator. 
As both EBMLE and the URE estimators (as well as the pre-test estimator) are designed for the additive model, the results from this analysis might be considered a better basis for comparison between the methods. 


\section{Discussion}\label{sec:discussion}
We considered estimation under sum-of-squares loss of the cell means in a two-way linear model with additive fixed effects, where the focus was on the unbalanced case. 
Minimax shrinkage estimators exist which differ from, and hence dominate, the Least Squares estimator for the more general linear regression setup \citep{rolph1976choosing}.
However, such estimators do not exploit the special structure of the two-factor additive model, and might lead to undesirable shrinkage patterns which are difficult to interpret. 
Instead, we considered a parametric class of Bayes estimators corresponding to a prior motivated from exchangeability considerations. 
The resulting estimates exhibit meaningful shrinkage patterns and, when appropriately calibrated, achieve significant risk reduction as compared to the Least Squares estimator in practical situations. 

To calibrate the Bayes estimator we considered substituting the hyperparameters governing the prior with data-dependent values, and proposed a method which chooses these values in an asymptotically optimal way. We contrasted the proposed estimator with the traditional likelihood based empirical BLUP estimator, which was shown to generally produce asymptotically sub-optimal estimates of the cell means. Since it relies on the postulated two-level model, the likelihood based empirical BLUP estimator might be led astray when there is dependency between the cell counts and the true cell means; this was clearly shown in our simulation examples. 

The theory developed here employs proof techniques that differ in fundamental aspects from those commonly used to prove asymptotic optimality in the one-way normal mean estimation problem. 
We offered a flexible approach for proving asymptotic optimality by showing efficient point-wise risk estimation.  It greatly helped to tackle the  difficulties encountered in two-way problem, where computations involving matrices are generally unavoidable. Our proof techniques can be extended to $k$-way additive models, although computational difficulty might become a problem when $k$ is even moderately large. It would be interesting to investigate whether computationally efficient methods can be developed for the higher-way unbalanced layout.  


\section{Acknowledgement}
The authors would like to thank Tony Cai, Samuel Kou and Art Owen for helpful discussions.

\appendix
\section{Proofs of the asymptotic optimality results of Section~\ref{sec:theory}}\label{append.theory}

Throughout this section we present our proofs assuming  $\sigma =1$. 
It is done mainly for the ease of presentation, and the proofs can easily be modified for any arbitrary but known value of $\sigma$. 
Next we introduce some notation. 
We denote by $\sigma_k(A)$ the $k$-th largest singular value of a matrix $A$. 
We denote by $\lambda_k(B)$ the $k$-th largest eigenvalue of a symmetric matrix $B$. Also, we denote $G \stackrel{\cdot}{=} M \Vinv$ and $H \stackrel{\cdot}{=} G^{\topp}QG = \Vinv M Q M \Vinv$ where $M$, $\Vinv$ and $Q$ are defined in Section~\ref{sec:missing}. We define  $W = M^{\frac{1}{2}} \Vinv M^{\frac{1}{2}}$.
As  $0 \prec M\preceq \V$ we have $W\preceq I$, and also $W^2\preceq I$. We will use the following result of \citealp{searle2009variance} (Theorem S4, Page 467).
\begin{lemma}{Central moments of Gaussian Quadratic Forms.}\label{var.lemma}
	If $\y \sim N(\etab,V)$ then: 
	 $$\ex(\y^{\topp} A \y) = 2 \tr[AV] + \etab^{\topp}A\etab~, \quad \text{ and } \quad \Var(\y^{\topp} A \y) = 2 \tr[(AV)^2] + 4 \etab^{\topp} AVA \etab~.$$
\end{lemma}

\subsection{Proof of Theorem~\ref{thm:loss-approx-missing}, Lemma~\ref{lem:risk-1} and Lemma~\ref{lem:risk-2}}
The proof of Theorem~\ref{thm:loss-approx-missing} for the case when the general effect $\mu=0$ follows  directly  from the results of Lemma~\ref{lem:risk-1} and Lemma~\ref{lem:risk-2} as $ \ex \{ \UREQ_{r,c}( 0, \lambdaa, \lambdab) -  L_{r,c}^Q\big(\etab, \etahatsb(0, \lambdaa, \lambdab)\big)\}^2$ is bounded above by
\begin{align*}
&  2 \, \ex \Big\{ \UREQ_{r, c}( 0, \lambdaa, \lambdab) - R^Q_{r,c}(\etab, \etahatsb(0, \lambdaa, \lambdab)) \Big\}^2 \qquad \qquad \qquad \qquad \qquad \qquad\qquad \qquad \qquad\\
& + \, 2 \, \ex \Big\{ L_{r, c}^Q\big(\etab, \etahatsb(0, \lambdaa, \lambdab)\big) - R^Q_{r,c}(\etab, \etahatsb(0, \lambdaa, \lambdab)) \Big\}^2 .
\end{align*}
In fact, in this case we actually prove Theorem~\ref{thm:loss-approx-missing} with the stronger $L_2$ norm. 
We now concentrate on proving the lemmas; we will prove the theorem for the general case later by building on the proofs for the $\mu=0$ case.\\[2ex]  
\textbf{Proof of  Lemma~\ref{lem:risk-1}.}
As the URE is an unbiased estimator of the risk of an estimator in $\mathcal{S}$,  for any fixed $\lambdaa, \lambdab \geq 0$, we have
\begin{align}\label{temp.1}
\ex[ \UREQ(0, \lambdaa, \lambdab) -  R^Q_{r,c}(\eta; \etahatsb(0, \lambdaa, \lambdab))]^2 &= \Var[ \UREQ(0, \lambdaa, \lambdab) ]
\end{align}
Based on the expression of the URE estimator in \eqref{eq:sure-missing} we know that
$$\UREQ (0, \lambdaa, \lambdab)  = (rc)^{-2} \big \{\sigma^2 \tr(QM)  - 2\sigma^2 \tr (\Vinv MQM) + \y^{\topp} H \y\big\},$$
and so the RHS of \eqref{temp.1} reduces to $(rc)^{-2} \Var(\y^{\topp}H\y)$  which, being the variance of a quadratic form of the Gaussian random vector $\y$, can in turn be evaluated by using Lemma~\ref{var.lemma} to give
\begin{align}\label{eq:var-sure-zero} 
 \Var[ \UREQ(0, \lambdaa, \lambdab) ] = (rc)^{-2} \{ 2 \tr(HMHM) + 4\etab^{\topp}HMH\etab \}.
\end{align} 
Our goal now is to show that each of the terms on the RHS, after being multiplied by $d_{r,c}^2$, uniformly converges to $0$ for all choices of  $\lambdaa$ and $\lambdab$. 
For this purpose, we concentrate on the second term of the RHS first. As $H$ is p.s.d. by \texttt{R\ref{fc:psd-butterfly}} (See Section~\ref{sec:lin_results} of Supplement),  $HMH$ is also p.s.d. Thus,  $\etab^{\topp} H M H \etab \leq \lambda_1(H M H) \|\etab\|^2$. Next, we bound the largest eigen value of $HMH$ as  
\begin{align*}
\lambda_1(HMH) &= \lambda_1(\Vinv M Q M \Vinv M \Vinv M Q M \Vinv) \\
& = \lambda_1(\Vinv M Q M^\half W^2 M^\half Q M \Vinv) \\
& \leq \lambda_1( \Vinv M Q M Q M \Vinv ).
\end{align*}
The last inequality uses $W^2\preceq I$. Again, by \texttt{R\ref{fc:interchange}} of Supplement Section~\ref{sec:lin_results}, the RHS above  equals $\lambda_1( M^\half Q M \Vinv \Vinv M Q M^\half )$. Thus, we have
\begin{align*}
\lambda_1(HMH) &= \lambda_1( M^\half Q M \Vinv \Vinv M Q M^\half )\\
&= \lambda_1( M^\half Q M^\half W \Minv W M^\half Q M^\half ) \\
&\leq \lambda_1(\Minv) \lambda_1(M^\half QMQ M^\half)\\
&=\lambda_1(\Minv) \lambda_1^2(M^\half Q M^\half). 
\end{align*}
The  inequality follows by using $W \Minv W \preceq \lambda_1(\Minv)I$. 
Thus, we arrive at the following upper bound 
 $$(rc)^{-2} d_{r,c}^2\, \sup_{\lambdaa,\lambdab \geq 0} \etab^{\topp} H M H \etab \leq (rc)^{-2} \,d_{r,c}^2\, \lambda_1(\Minv) \lambda_1^2(M^\half Q M^\half) \|\etab\|^2$$
which, under Assumptions A1 and A2, converges to $0$ as $r,c \to \infty$. 
Now, for the first term in \eqref{temp.1} we have
\begin{align*}
\tr(HMHM) 
&= \tr(M^\half H M H M^\half)\\ 
&= \tr(M^\half \Vinv MQM \Vinv M \Vinv MQM \Vinv M^\half) \\
&=\tr( W M^\half Q M^\half W^2 M^\half Q M^\half W ) \\
& \leq \tr( W M^\half Q M Q M^\half W ) \qquad \qquad  &&[\text{using } W^2 \preceq I]\\
&= \tr ( M^\half Q M^\half W^2 M^\half Q M^\half )  \qquad  \qquad &&[\text{we use \texttt{R\ref{fc:interchange}} here}]\\
&\leq \tr( M^\half QMQ M^\half ) \qquad  \qquad &&[\text{again using } W^2 \preceq I]\\
& \leq (rc) \cdot \lambda_1( M^\half QMQ M^\half ) \\
&=  (rc) \cdot \lambda_1^2( M^\half Q M^\half ),
\end{align*}
where the last equation follows by using \texttt{R\ref{fc:interchange}} again. Thus,  as $r,c \to \infty$, by Assumption A2 the first term in \eqref{temp.1} scaled by $d_{r,c}^2$ also converges to $0$ uniformly over the ranges of $\lambdaa$ and $\lambdab$. 
This completes the proof of the lemma.\medskip\\

\noindent \textbf{Proof of Lemma~\ref{lem:risk-2}}
As the risk is the expectation of the loss, to prove the lemma we need to show: 
\bes
d_{r,c}^2 \, \sup_{\lambdaa,\lambdab \geq 0} \Var \big[L^Q(\etab, \etahatsb(0, \lambdaa, \lambdab))  \big] \to 0 \text{ as } r,c\to \infty. 
\ees
Again, the loss of the estimator $\etahatsb_0 = \etahatsb(0, \lambdaa, \lambdab)$ can be decomposed as 
\begin{align*}
L^Q(\etab, \etahatsb_0) &= (rc)^{-1} (\etahatsb_0 - \etab)^{\topp} Q (\etahatsb_0 - \etab) 
= (rc)^{-1} (\y - \etab - G\y)^{\topp} Q (\y - \etab - G\y) \\
&= (rc)^{-1} \big\{ (\y-\etab)^{\topp} Q (\y-\etab) + \y^{\topp}H\y - 2(\y-\etab)^{\topp}QG\y \big\} \\
&= (rc)^{-1} \big\{L_1 + L_2 - L_3 + L_4 \big\},
\end{align*}
where $L_1=(\y-\etab)^{\topp} Q (\y-\etab), \;\; \;
L_2=\y^{\topp} H \y,\;\; \;
L_3=2\y^{\topp} QG \y,\;\; \;
L_4=2\etab^{\topp} QG \y.
$\\[1ex]
Hence, it suffices to show that $ d_{r,c}^2\,\sup_{\lambdaa, \lambdab}\Var((rc)^{-1} L_i) \to 0$ as $r,c \to \infty$ for all $i=1,\ldots,4$. 
Uniform convergence of the desired scaled variance of $L_2$ was already shown in the proof of Lemma~\ref{lem:risk-1}.\\[1ex]
For the first term $L_1$ we have:
\begin{align*}
\Var[(rc)^{-1} L_1] &= (rc)^{-2} \Var[(\y-\etab)^{\topp} Q (\y-\etab)] = (rc)^{-1} 2 \,\tr(QMQM)\\
&=(rc)^{-2} 2 \, \tr\{(M^\half Q M^\half)^2\} \leq (rc)^{-1} 2 \, \lambda_1^2(M^\half Q M^\half)
\end{align*} 
which by Assumption A2 
is $o(d_{r,c}^{-2})$ as $r,c \to \infty$ for any value of the hyper-parameter.  
As $\y$ is normally distributed, the forth term can be explicitly evaluated as  
\begin{align*}
4^{-1}\Var(L_4)=\Var(\etab^{\topp} QGy) 
= \etab^{\topp} QGMG^{\topp}Q\etab
&\leq \lambda_1(QGMG^{\topp}Q) \|\etab\|^2 \\
& = \lambda_1(QM\Vinv M\Vinv MQ) \|\etab\|^2 \\
&= \lambda_1(QM^\half W^2M^\half Q) \|\etab\|^2 \\
& \leq \lambda_1(Q M^\half M^\half Q) \|\etab\|^2 \\
&\leq \lambda_1(M^{-1})\lambda_1^2(M^\half QM^\half) \|\etab\|^2
\end{align*}
which by assumptions A1-A2 is 
$o(r^2 c^2 d_{r,c}^{-2})$.\\
The third term requires detailed analysis. 
First, note that it breaks into two components
\begin{align}\label{temp.2}
\Var[(rc)^{-1}L_3]&= 4 (rc)^{-2} \Var(\y^{\topp} Q G \y) \\
& = 8 (rc)^{-2} \tr(\Gtil M \Gtil M) + 16 (rc)^{-2} \etab^{\topp} \Gtil M \Gtil \etab 
\end{align}
$\text{ where, } \Gtil = QG + G^{\topp}Q$ is a symmetric matrix.  
We concentrate on the second term of the RHS first. 
Note that 
$$ (rc)^{-2} \etab^{\topp} \Gtil M \Gtil \etab  \leq  (rc)^{-2} \etab^{\topp} \etab \;  \sigma_1(\Gtil M \Gtil)~.$$
Like before, if we can uniformly bound the largest eigen value of $\Gtil M \Gtil$ as $o(rc d_{r,c}^{-1})$ then the above is $o(d_{r,c}^{-2})$ as $ r, c \to \infty$ by Assumption A1.
Noting that $\Gtil = QG + G^{\topp}Q = QM\Vinv + \Vinv MQ$, we decompose
\begin{align*}
&\Gtil M \Gtil = H_1 + H_1^{\topp} + H_2 + H_3, \text{ where } 
H_1= QM\Vinv M Q M \Vinv, \; \\[1ex]
&H_2 = Q M\Vinv M \Vinv MQ\,, \; H_3= \Vinv MQMQM \Vinv.
\end{align*}
To uniform bound the eigen values of $\Gtil M \Gtil$ we just show that for each of $i=1,...,3$, $(rc)^{-1} \, d_{r,c}^{-2}\, \sigma_1(H_i) \to 0$ as $r,c \to \infty$.
$H_1$ is not a symmetric matrix. 
In this case, note that:
\begin{align*}
\sigma_1(H_1)&=\sigma_1(QM^\half W M^\half Q M^\half W M^{-\half})\\ 
&=\sigma_1(M^{-\half} M^\half QM^\half W M^\half Q M^\half W M^{-\half})\\
&\leq \lambda_1(M^{-\half} ) \cdot \sigma_1(M^\half QM^\half W M^\half Q M^\half W) \cdot \lambda_1(M^{-\half} )\\
&\leq \lambda_1(M^{-\half} ) \cdot \lambda_1(M^\half QM^\half) \cdot \lambda_1(W) \cdot \lambda_1(M^\half Q M^\half) \cdot \lambda_1(W) \cdot \lambda_1(M^{-\half})\\
&\leq  \lambda_1(M^{-1}) \cdot \lambda_1^2(M^\half QM^\half) 
\end{align*}
where the last inequality uses $W \preceq I$. 
For the symmetric matrix $H_2$ using $ W^2 \preceq I$, we have
\begin{align*}
\lambda_1(H_2) = \lambda_1(QM^\half W^2 M^\half Q) \leq \lambda_1(Q M^\half M^\half Q) \leq \lambda_1(M^{-1})\lambda^2_1(M^\half QM^\half)
\end{align*}
which is uniformly controlled at $o(r\,c \,d_{r,c}^{-2})$ by assumption A2. 
For the other symmetric matrix $H_3$ we also have
\begin{align*}
\lambda_1(H_3) 
&= \lambda_1(M^\half QM \Vinv \Vinv MQM^\half) \\
&= \lambda_1(M^\half QM^\half M^\half \Vinv M^\half \Minv M^\half \Vinv M^\half M^\half QM^\half) \\
&=\lambda_1(M^\half Q M^\half W \Minv W M^\half QM^\half) \\
&\leq \lambda_1(M^\half QM^\half) \lambda_1(W) \lambda_1(\Minv) \lambda_1(W) \lambda_1(M^\half QM^\half)\\
& \leq \lambda_1^2(M^\half QM^\half) \lambda_1(\Minv)
\end{align*}
which again is uniformly controlled at $o(rc\, d_{r,c}^{-2})$ by assumption A2.

Now we return to the first term in \eqref{temp.2} and upper bound $\tr(\Gtil M \Gtil M)$ by $o(r^2c^2\,d_{r,c}^{-2})$ when $r,c \to \infty$. Denote $\Gdot = QG$ so that $\Gtil = \Gdot + \Gdot^{\topp}$. 
We have
\bes\label{temp.3}
\tr(\Gtil M \Gtil M) = \tr(\Gdot M \Gdot M) + \tr(\Gdot^{\topp} M \Gdot^{\topp} M) +  2 \tr(\Gdot M \Gdot^{\topp} M). 
\ees
Substituting the expression of $\Gtil$ we get
\bes
\Gdot M \Gdot M 
= QM\Vinv M QM\Vinv M
= QM^\half W M^\half Q M^\half Q M^\half W M^\half,
\ees
and so we can upper bound its trace as
\begin{align*}
\tr(\Gdot M \Gdot M) 
&= \tr(W M^\half Q M^\half W M^\half Q M^\half) 
\leq \lambda_1(W) \tr(M^\half Q M^\half W M^\half Q M^\half)\\
&\leq \tr( M^\half Q M^\half M^\half Q M^\half ) 
\leq rc \cdot \lambda_1^2(M^\half Q M^\half)
=o(r^2c^2\, d_{r,c}^{-2})
\end{align*}
for any $\Vinv$ and any $M, Q$ which obeys Assumption A2. Noting that $\tr(\Gdot^{\topp} M \Gdot^{\topp} M) = \tr(M \Gdot M \Gdot) = \tr(\Gdot M \Gdot M)$, the second term in \eqref{temp.3} is also uniformly bounded by $o(r^2c^2\,d_{r,c}^{-2})$. Finally, for the third term we have 
\bes
\Gdot M\Gdot^{\topp} M 
= QM\Vinv M \Vinv M Q M 
=QM^\half W^2 M^\half Q M 
\preceq QMQM,
\ees
and so its trace is upper bounded by
\begin{align*}
\tr(\Gdot M\Gdot^{\topp} M) 
&\leq \tr(QMQM) = \tr(M^\half Q M^\half)^2 =rc\cdot \lambda_1^2(M^\half Q M^\half)
=o(r^2c^2\,d_{r,c}^{-2})
\end{align*}
by Assumption A2. Thus, we conclude that $\tr(\Gtil M \Gtil M)$ is uniformly bounded by $o(r^2c^2\,d_{r,c}^{-2})$ as $r,c \to \infty$. This complete the proof of the lemma.\medskip\\ 

\noindent \textbf{Proof of Theorem~\ref{thm:loss-approx-missing} for the general case.}
Using the above two lemmas, we now prove our main theorem for the general case. 
First, note that
for arbitrary fixed $\mu \in \R$, the loss function decomposes into the following components: 
\begin{align*}
(\etahatsb(\mu, \lambdaa, \lambdab) - \etab)^{\topp} Q (\etahatsb(\mu, \lambdaa, \lambdab) - \etab) &= (\etahatsb(0, \lambdaa, \lambdab) - \etab)^{\topp} Q (\etahatsb(0, \lambdaa, \lambdab) - \etab) \\
& + \mu^2 \one^{\topp} H \one -2\mu\one^{\topp}H\y + 2\mu\one^{\topp}G^{\topp}Q (\y-\etab).
\end{align*}
Comparing it with the definition of $\URE$ we have:
\begin{align*}
\UREQ_{r,c}( \mu, \lambdaa, \lambdab) -  L_{r,c}^Q\big(\etab, \etahatsb(0, \lambdaa, \lambdab)\big)  = & 
\UREQ_{r,c}( 0, \lambdaa, \lambdab) -  L_{r,c}^Q\big(\etab, \etahatsb(0, \lambdaa, \lambdab)\big) \\ & + 2 (rc)^{-1}\mu\one^{\topp}G^{\topp}Q (\y-\etab).
\end{align*}
We have already proved the theorem for the case of $\mu=0$; hence, in light of the above identity, the proof of the general case will follow if we can show:
\begin{align}\label{eq:temp.4}
 \lim_{\substack{\hspace{1mm}\\ r \to \infty \\[0.5ex] c \to \infty}} \;\;  \sup_{\substack{|\mu| \leq \thr\\[0.5ex] \lambdaa, \lambdab\geq 0}}  d_{r,c} \cdot (rc)^{-1} \cdot \ex \big \vert \mu\one^{\topp}G^{\topp}Q (\y-\etab) \big \vert = 0~. 
 \end{align}
Noting that for any fixed $\eta$ the random variable $F=\one^{\topp}G^{\topp}Q (\y-\etab)$ follows a univariate normal distribution with mean $0$ and variance $\one^{\topp}G^{\topp}QMQ G \one$, the above holds if
\begin{align}\label{eq:temp.4}
\lim_{ r \to \infty, c \to \infty} \;\; \thr \, \cdot\, d_{r,c} \cdot (rc)^{-1} \,  \cdot  \sup_{\lambdaa, \lambdab\geq 0}  \big\{\Var \big( \big \vert \one^{\topp}G^{\topp}Q (\y-\etab) \big \vert\big)\big\}^{1/2} = 0~. 
\end{align}
Bounding the variance of $F$ as
\begin{align*}
\Var(\one^{\topp} G^{\topp}Q\y) 
&\leq \mu^2\one^{\topp}G^{\topp}QMQG\one\\
&=\one^{\topp}\Vinv M Q M Q M \Vinv\one\\
&\leq \one^{\topp}\one \cdot \lambda_1(\Vinv M^\half M^\half Q M^\half M^\half Q M^\half M^\half \Vinv) \\
&=  rc \cdot \lambda_1(M^\half Q M^\half M^\half \Vinv \Vinv M^\half M^\half Q M^\half)\\
&\leq  rc \cdot \lambda_1( M^\half Q M^\half )  \lambda_1(M^\half \Vinv \Vinv M^\half)  \lambda_1( M^\half Q M^\half )\\
&=  rc \cdot \lambda_1( M^\half Q M^\half )  \lambda_1(WM^{-1}W)  \lambda_1( M^\half Q M^\half )\\
&\leq  rc \cdot \lambda_1^2 ( M^\half Q M^\half ) \lambda_1(M^{-1}) \leq  rc \cdot \lambda_1^2 ( M^\half Q M^\half ) \lambda_1(M^{-1}),
\end{align*}
\eqref{eq:temp.4} is proved. 

\subsection{Proof of the Decision Theoretic results: Theorems~\ref{thm:asym-ol-pr-missing}, \ref{thm:asym-ol-ev-missing} and Corollary~\ref{cor:asym-opt-missing}}\label{append:theory-2}
\textbf{Discretization.} In this section, we first define analogous versions of the URE and oracle estimators over a discrete set. Note that in \eqref{eq:sure-def-missing} and \eqref{eq:ol-missing} the URE and oracle estimators involve minimizing the hyper-parameters $(\mu, \lambdaa, \lambdab)$ simultaneously over $\hat{T}_{r,c}=[\hat{a}_{\tau},   \hat{b}_{\tau}] \times [0, \infty] \times [0,\infty]$ where the range of the location hyper-parameter $\mu$ depends on the data. 
We define a discrete product grid $\net=\net^{[1]} \times \net^{[2]} \times \net^{[3]}$ 
which only depends on $r,c$ and not on the data. Details for the
construction of $\net$ is provided afterwards. It contains countably infinite grid points as $r, c \to \infty$.
We define the discretized version of the  oracle estimator where the minimization is conducted over all the points in the discrete grid $\net$ that are contained in  $\hat{T}_{r,c}$. We define the discretized oracle loss hyper-parameters as
\bes
\big( \ \mutil^{\rm OD}, \lambdaatil^{\rm OD}, \lambdabtil^{\rm OD} \ \big) = \argmin_{(\mu,\lambdaa,\lambdab) \, \in \, \net \, \cap \, \hat{T}_{r,c}}  L^Q_{r,c}\big(\etab, \etahatsb(\mu, \lambdaa, \lambdab)\big) ~,
\ees 
and the corresponding oracle rule by
$\etatb_c^{\rm OD} = \Ztil \Zinv \etahatsb( \mutil^{\rm OD}, \lambdaatil^{\rm OD}, \lambdabtil^{\rm OD} )$. We define the URE estimators over the discrete grid by projecting the URE estimates of equation~\eqref{eq:sure-def-missing} in $\nett_{r,c} \cap \hat{T}_{r,c}$: 
if the URE hyper-parameters given by Equation~\eqref{eq:sure-def-missing} are such that:
$$\mu_1 \leq \muhat^{\rm U_Q} \leq \mu_2\,, \quad  \lambda_1 \leq \lambdaahat^{\rm U_Q} \leq \lambda_2\,, \quad \text{ and } \quad \lambda_3 \leq \lambdabhat^{\rm U_Q} \leq \lambda_4 $$
where $\mu_1, \mu_2$ are neighboring points in $\net^{[1]} \cap [\hat{a}_{\tau}, \hat{b}_{\tau}]$,    $\lambda_1, \lambda_2$ are neighboring points in   $\net^{[2]}$ and $\lambda_3, \lambda_4$ are neighboring points in  $\net^{[3]}$, then 
the URE estimates of the tuning parameters over the discrete grid is defined as the minima over the nearest $8$-point subset of the grid: 
\begin{align}\label{eq:UD}
(\muhat^{\rm UD}, \lambdaahat^{\rm UD}, \lambdabhat^{\rm UD}) = \argmin_{ (\mu,\lambdaa,\lambdab) \in \{\mu_1,\mu_2\} \times \{\lambda_1,\lambda_2\} \times \{\lambda_3,\lambda_4\} }  \UREQ\big(\mu, \lambdaa, \lambdab\big). 
\end{align}
The corresponding discretized EB estimate is $\etahatb^{\rm UD}=\etahatsb(\muhat^{\rm UD}, \lambdaahat^{\rm UD}, \lambdabhat^{\rm UD})$. 
The corresponding estimate for $\etatil$ is 
$ \etahatb_{\sf c}^{\rm UD} = \Ztil \Zinv \etahatsb(\muhat^{\rm D}, \lambdaahat^{\rm D}, \lambdabhat^{\rm D})$.
If the URE estimators for any of the three hyper-parameters are outside the grid then the nearest boundary of the grid is taken as the $\rm{UD}$ estimate for that hyper-paramter. We will show afterwards that the probability of such events is negligible. 
Note that, by construction, $L(\etab_c, \etahatb_c^{\rm UD}) \geq L(\etab_c, \etatb_c^{\rm OD})\geq L(\etab_c, \etatb_c^{\rm OL})$. \\[1ex]
\par
\noindent \textbf{Construction of the grid $\nett_{r,c}$.} The grid $\nett_{r,c}$ is a product grid. The grid $\net^{[1]}$  on the location hyper-parameter $\mu$ is an equispaced discrete set $\{-m_{r,c}=\mu[1]<\mu[2]<\cdots<\mu[n_1]\leq m_{r,c} \}$ which  covers $[-\thr,\thr]$ at a spacing of $\delta_{r,c}^{[1]}$. Thus, the cardinality of $\net^{[1]}$, $n_1= \lceil 2 \thr \{\delta_{r,c}^{[1]}\}^{-1} \rceil$. We choose the spacing as
\begin{align}\label{eq.delta1}
\delta_{r,c}^{[1]}= \{\thr^{4/3} \cdot \nu_{r,c} \cdot \lambda_1(Q) \}^{-1}~.
\end{align}
 For constructing the grid $\net^{[2]}$ on the scale hyper-parameter,  we consider the following transformation $\lambdaatil=(1+\lambdaa)^{-1/2}$. Note that $\lambdaatil \in [0,1]$ as $\lambdaa$ varies over $[0,\infty]$. We construct an equispaced grid on $\lambdaatil$ between $0$ and $1$ at a spacing of  $\delta_{r,c}^{[2]}$:
 $$\{0=\lambdaatil[1]<\lambdaatil[2]<\cdots<\lambdaatil[n_2]\leq 1 \}\, \text{ where } \lambdaatil[k]=(k-1)\delta_{r,c}^{[2]} \text{ and }  n_2= \lceil  \{\delta_{r,c}^{[2]}\}^{-1} \rceil.$$
 The grid on $\lambdaatil$ is then retransformed to produce the grid $\net^{[2]}$ on the scale hyper-parameter $\lambdaa$ in the domain $[0,\infty]$. 
 The grid $\net^{[3]}$ on $\lambdab$ is similarly constructed with $\delta_{r,c}^{[3]}$ distances between two corresponding grid points in $\lambdabtil$ scale. The spaces were chosen as:  
\begin{align}\label{eq.delta2}
\delta_{r,c}^{[2]}=\delta_{r,c}^{[3]}= \{\thr^{7/3} \cdot \nu_{r,c} \cdot \lambda_1(Q) \}^{-1}~.
\end{align}
Now, as $r, c \to \infty$, $n_1= O(\thr^{7/3} \cdot \nu_{r,c} \cdot \lambda_1(Q))$, $n_2=O(\thr^{7/3} \cdot \nu_{r,c} \cdot \lambda_1(Q))$ and thus the cardinality of $\net$ is $|\net| = O(\thr^7\, \nu_{r,c}^3\, \lambda_1^3(Q))=O(d_{r,c})$.
\par
The following two lemmas enable us to work with the more tractable, discretized versions of the URE and oracle estimators when proving the decision theoretic results. 
The first one shows that the difference in the loss between the true estimators and their discretized versions  is asymptotically controlled at any prefixed level. The second shows that the URE values for the estimator is also asymptotically close for the discretized version. 

\begin{lemma}\label{lem:sec4.temp}
	For any fixed $\epsilon > 0$, under Assumptions A1-A2, 
	\begin{align*}
		\texttt{A.} \quad & P\big\{ L_{r, c} (\etatil, \etatb_{\sf c}^{\rm OD}) -  L_{r, c} (\etatil, \etatb_{\sf c}^{\rm OL}) > \epsilon \big\} \to 0 \text{ as } r, c \to \infty	 \;\; \text{ and },\\
		\texttt{B.} \quad  & \ex |L_{r, c} (\etatil, \etatb_{\sf c}^{\rm OD}) -  L_{r, c} (\etatil, \etatb_{\sf c}^{\rm OL})| \to 0 \text{ as } r, c \to \infty~,\\
	\texttt{C.} \quad & P\big\{  \big \vert L_{r, c} (\etatil, \etahatb_{\sf c}^{\rm UD}) -  L_{r, c} (\etatil, \etahatb_{\sf c}^{\rm URE}) \big \vert > \epsilon \big\} \to 0 \text{ as } r, c \to \infty~,	 \;\;\\
	\texttt{D.} \quad  & \ex |L_{r, c} (\etatil, \etahatb_{\sf c}^{\rm UD}) -  L_{r, c} (\etatil, \etahatb_{\sf c}^{\rm URE})| \to 0 \text{ as } r, c \to \infty~. 
	\end{align*}
\end{lemma}

\begin{lemma}\label{lem:sec4.temp.final}
	For any fixed $\epsilon > 0$, under Assumptions A1-A2, 
	\begin{align*}
	\texttt{A.} \quad & P\big\{  \, \UREQ( \muhat^{\rm UD}, \lambdaahat^{\rm UD}, \lambdabhat^{\rm UD}) -  \UREQ( \muhat^{\rm U_Q}, \lambdaahat^{\rm U_Q}, \lambdabhat^{\rm U_Q}) > \epsilon \big\} \to 0 \text{ as } r, c \to \infty	 \;\; \text{ and},\\
	\texttt{B.} \quad  & \ex \big [ \UREQ( \muhat^{\rm UD}, \lambdaahat^{\rm UD}, \lambdabhat^{\rm UD}) -  \UREQ( \muhat^{\rm U_Q}, \lambdaahat^{\rm U_Q}, \lambdabhat^{\rm U_Q})\big ]\to 0 \text{ as } r, c \to \infty~. 
	\end{align*}
\end{lemma}

\noindent The proof of Lemma~\ref{lem:sec4.temp} uses the following two lemmas. For shortage of space,  the proofs of all these other lemmas (\ref{lem:sec4.temp}, \ref{lem:sec4.temp.final}, \ref{lem:sec4.temp.1} and \ref{lem:sec4.temp.2}) is provided in the supplementary materials. 

\begin{lemma}\label{lem:sec4.temp.1}
	Under assumption A1 on the parametric space, for any fixed $\tau \in (0,1]$, $\thr=\log (rc)$, the event $A_{r,c}(\Y)=\{[\hat{a}_{\tau}, \hat{b}_{\tau}] \subseteq [-\thr,\thr]\}$ satisfies 
	$$P\big\{ A_{r,c} \big\} \to 1 \text{ as } n \to \infty~.$$
\end{lemma}

\begin{lemma}\label{lem:sec4.temp.2}
	Under assumptions A1-A2, for any fixed $\tau \in (0,1]$, $\thr=\log (rc)$, the event $A_{r,c}(\Y)=\{[\hat{a}_{\tau}, \hat{b}_{\tau}] \subseteq [-\thr,\thr]\}$ satisfies:
	\begin{align*}
	\texttt{A.} \quad  & \ex \big\{|L_{r, c} (\etatil, \etatb_{\sf c}^{\rm OD}) -  L_{r, c} (\etatil, \etatb_{\sf c}^{\rm OL})| \cdot I\{A_{r,c}(\Y)\} \big\} \to 0 \text{ as } n \to \infty.\\
	\texttt{B.} \quad  & \ex \{|L_{r, c} (\etatil, \etahatb_{\sf c}^{\rm UD}) -  L_{r, c} (\etatil, \etahatb_{\sf c}^{\rm URE})|\cdot I\{A_{r,c}(\Y)\} \to 0 \text{ as } r, c \to \infty .
	\end{align*}
\end{lemma}

We next present the proof of the decision theoretic properties where Lemmas~\ref{lem:sec4.temp}, \ref{lem:sec4.temp.final} will be repeatedly used.\\

\noindent \textbf{Proof of Theorem~\ref{thm:asym-ol-pr-missing}}. 
We know that
\begin{align*}
 P \big\{  L(\etab_c, \etahatb_c^{\rm URE}) \geq L(\etab_c, \etatb_c^{\rm OL}) + \epsilon \big\} \leq & P \big\{  L(\etab_c, \etahatb_c^{\rm URE}) \geq L(\etab_c, \etatiltil^{\rm OD}) + \epsilon/2 \big\}  \\
& + P \big\{  L(\etab_c, \etatiltil^{\rm OD} \geq L(\etab_c, \etatb_c^{\rm OL}) + \epsilon/2 \big\}. 
\end{align*}
The second term converges to $0$ by Lemma~\ref{lem:sec4.temp}. The first term is again less than:
\begin{align*}
P \big\{  L(\etab_c, \etahatb_c^{\rm UD}) \geq L(\etab, \etatiltil^{\rm OD}) + \epsilon/4 \big\} +  P \big\{  |L(\etab_c, \etahatb_c^{\rm URE}) - L(\etab, \etahatb_c^{\rm UD})| \leq \epsilon/4 \big\} \,.
\end{align*}
The second term in the RHS above converges to $0$ as $r,c \to \infty$ by Lemma~\ref{lem:sec4.temp}. For the first term note that, by definition, $\UREQ( \muhat^{\rm U_Q}, \lambdaahat^{\rm U_Q}, \lambdabhat^{\rm U_Q}) \leq \UREQ( \mutil^{\rm OD}, \lambdaatil^{\rm OD}, \lambdabtil^{\rm OD})$ which, combined with Lemma~\ref{lem:sec4.temp.final}, suggests that $$P\big\{\;\UREQ( \muhat^{\rm UD}, \lambdaahat^{\rm UD}, \lambdabhat^{\rm UD}) \leq \UREQ(\mutil^{\rm OD}, \lambdaatil^{\rm OD}, \lambdabtil^{\rm OD})+\epsilon/8\;\big\}\to 0 \text{ as } r, c \to \infty~.$$
Thus, showing $P \big\{  L(\etab_c, \etahatb_c^{\rm UD}) \geq L(\etab, \etatiltil^{\rm OD}) + \epsilon/4 \big\}  \to 0$ as $r,c \to \infty$ can be reduced to showing the following:
\begin{equation*}
\lim_{r,c \to \infty } P \big\{  A(\y; \etab_c) \geq B(\y; \etab_c) + \epsilon/8 \big\} = 0 
\end{equation*}
where
\begin{align*}
 A(\y; \etab_c) &= L(\etab_c, \etahatb_c^{\rm UD}) - \UREQ( \muhat^{\rm UD}, \lambdaahat^{\rm UD}, \lambdabhat^{\rm UD})\\
 B(\y; \etab_c) &= L(\etab_c, \etatiltil^{\rm OD}) - \UREQ( \mutil^{\rm OD}, \lambdaatil^{\rm OD}, \lambdabtil^{\rm OD}).
\end{align*}
Noting that $L(\etab_c, \etahatb_c^{\rm UD}) = L^Q(\etab, \etahatb^{\rm UD})$ and $L(\etab_c, \etatiltil^{\rm OD})=L^Q(\etab, \etahatsb(\mutil^{\rm OD}, \lambdaatil^{\rm OD}, \lambdabtil^{\rm OD}))$, by using Markov's inequality we get 
\begin{align*}
P\big\{  A(\y; \etab_c) \geq B(\y; \etab_c) + \epsilon/8 \big\} &\leq 8^{-1}\epsilon^{-1} \ex \{| A(\y; \etab_c) - B(\y; \etab_c) |\}.
\end{align*}
By Triangle inequality the RHS above is upper bounded by
\begin{align*}
 & 16 \, \epsilon^{-1}  \ex \bigg\{\sup_{(\mu,\lambdaa,\lambdab) \in \nett_{r,c}}| L^Q(\etab, \etahatsb(\mu, \lambdaa, \lambdab)) - \UREQ( \mu,\lambdaa, \lambdab) | \bigg \}\\
 & \leq 16 \, \epsilon^{-1}  \ex \bigg\{\sum_{(\mu,\lambdaa,\lambdab) \in \nett_{r,c}}| L^Q(\etab, \etahatsb(\mu, \lambdaa, \lambdab)) - \UREQ( \mu,\lambdaa, \lambdab) | \bigg \}\\
 & \leq 16 \, \epsilon^{-1} |\nett_{r,c}| \sup_{\substack{|\mu| \leq \thr \\[0.5ex] \lambdaa, \lambdab\geq 0}} \ex \bigg\{| L^Q(\etab, \etahatsb(\mu, \lambdaa, \lambdab)) - \UREQ( \mu,\lambdaa, \lambdab) | \bigg \}.
 \end{align*}
As $|\nett_{r,c}| = O(d_{r,c})$ by Theorem~\ref{thm:loss-approx-missing}, the above expression converges to zero when $r,c \to \infty$. This completes the proof of the theorem.\\[1ex]

\noindent \textbf{Proof of Theorem~\ref{thm:asym-ol-ev-missing}}.
We decompose the loss into the following three components:
\begin{align*}
\{L(\etab_c, \etahatb_c^{\rm URE}) - L(\etab_c, \etahatb_c^{\rm UD})\}+\{ L(\etab_c, \etatb_c^{\rm OD})- L(\etab_c, \etatb_c^{\rm OL})\} + \{L(\etab_c, \etahatb_c^{\rm UD})-L(\etab_c, \etatb_c^{\rm OD}) \}.
\end{align*}
By Lemma~\ref{lem:sec4.temp}, the expectation of the absolute value of the first two terms converges to $0$ as $r,c \to \infty$. The third term is further decomposed as
\begin{align*}
L(\etab_c, \etahatb_c^{\rm UD}) - L(\etab_c, \etatb_c^{\rm OD}) =& \{ L(\etab_c, \etahatb_c^{\rm UD}) - \UREQ(\muhat^{\rm UD}, \lambdaahat^{\rm UD}, \lambdabhat^{\rm UD}) \} \\
& - \{ L(\etab_c, \etatb_c^{\rm OD}) - \UREQ(\mutil^{\rm OD}, \lambdaatil^{\rm OD}, \lambdabtil^{\rm OD}) \} \\
& + \{ \UREQ( \muhat^{\rm UD}, \lambdaahat^{\rm UD}, \lambdabhat^{\rm UD}) - \UREQ( \mutil^{\rm OD}, \lambdaatil^{\rm OD}, \lambdabtil^{\rm OD}) \}. 
\end{align*}
By definition $\UREQ( \muhat^{\rm U_Q}, \lambdaahat^{\rm U_Q}, \lambdabhat^{\rm U_Q}) \leq \UREQ( \mutil^{\rm OD}, \lambdaatil^{\rm OD}, \lambdabtil^{\rm OD})$ which, combined with Lemma~\ref{lem:sec4.temp.final}, suggests that the last term has asymptotically non-positive expectation.  Therefore, for all large $r, c$ values:
\begin{align*}
&\ex \big \{ L(\etab_c, \etahatb_c^{\rm UD}) - L(\etab_c, \etatb_c^{\rm OD}) \big \} \\
& \leq 2  \, \ex  \bigg\{\sup_{(\mu,\lambdaa,\lambdab) \in \nett_{r,c}} | L^Q(\etab, \etahatsb(\mu, \lambdaa, \lambdab)) - \UREQ(\mu, \lambdaa, \lambdab) | \bigg\}\\
& \leq 2  \, \ex  \bigg\{\sum_{(\mu,\lambdaa,\lambdab) \in \nett_{r,c}} | L^Q(\etab, \etahatsb(\mu, \lambdaa, \lambdab)) - \UREQ(\mu, \lambdaa, \lambdab) | \bigg\}\\
& \leq 2  \, |\nett_{r,c}| \sup_{ |\mu| \in \thr;  \lambdaa, \lambdab \geq 0} \ex  \bigg\{ | L^Q(\etab, \etahatsb(\mu, \lambdaa, \lambdab)) - \UREQ(\mu, \lambdaa, \lambdab) | \bigg\}. 
\end{align*}
As $|\nett_{r,c}|=O(d_{r,c})$, the above expression tends to zero when $r,c \to \infty$ by Theorem~\ref{thm:loss-approx-missing}. This completes the proof of Theorem~\ref{thm:asym-ol-ev-missing}. \\[1ex]

\noindent \textbf{Proof of Corollary~\ref{cor:asym-opt-missing}}. (a) and (b) are direct consequences, respectively, of Theorems ~\ref{thm:asym-ol-pr-missing} and \ref{thm:asym-ol-ev-missing}, since $L^Q(\etab, \etahatsb(\muhat, \lambdaahat, \lambdabhat)) \geq L^Q(\etab, \etab^{\rm OL})$ and, hence, also $\ex \{L^Q(\etab, \etahatsb(\muhat, \lambdaahat, \lambdabhat)) \} \geq \ex \{ L^Q(\etab, \etab^{\rm OL})\}$. 
Unlike in the above two theorems, here we only have optimality over the loss $L^Q$ defined over the observed cells with $Q$ in \eqref{eq:qmat}. 
As explained in Section \ref{sec:missing}, the loss $L^Q$ for the observed cells is the same as the (normalized) sum-of-squares loss over all (observed and missing) $rc$ cell means for an estimator of the form
$ \Ztil \Zinv \etahatsb(\muhat, \lambdaahat, \lambdabhat)$, 
where $\muhat, \lambdaahat, \lambdabhat$ are any estimates of the hyper-parameters.\\[1ex]
\noindent We end this section by proving the following interesting property of the $Q$ matrix.

\begin{lemma}\label{lem.assump.1}
	For $Q$ defined in \eqref{eq:loss.q1} we have $\lambda_1(Q) = \lambda_1\big((Z_c^TZ_c)(Z^TZ)^{\dagger}\big)$. Also, $\lambda_1(Q) \geq 1$ and 
	$\lambda_1(Q) =1$ if $Z=Z_c$.
\end{lemma}

\noindent \textbf{Proof of Lemma~\ref{lem.assump.1}.} By definition \eqref{eq:loss.q1}  we have 
\begin{align*}
\lambda_1(Q)=\lambda_1((Z_cZ^{\dagger})^TZ_cZ^{\dagger})=\lambda_1(Z_cZ^{\dagger}(Z_cZ^{\dagger})^T)=\lambda_1(Z_c(Z^TZ)^{\dagger}Z_c^{\topp})
\end{align*}
where the last equality follows as $Z^{\dagger}=(Z^TZ)^{\dagger}Z^{\topp}$ and so $(Z^{\dagger})^TZ^{\dagger}=(Z^TZ)^{\dagger}$. Thus we have $\lambda_1(Q) = \lambda_1\big((Z_c^TZ_c)(Z^TZ)^{\dagger}\big)$.

If $Z=Z_c$, then $\lambda_1(Q)=\lambda_1((Z_c^TZ_c)(Z_c^TZ_c)^{\dagger}\big)=1$ by definition of Moore-Penrose inverse. 
We will prove by contradiction that $\lambda_1(Q) \geq 1$ for any $Q$ under which $\etab$ is estimable. If possible assume $\lambda_1(Q) < 1$ which would imply $(Z_c^TZ_c)^{1/2} (Z^TZ)^{\dagger} (Z_c^TZ_c)^{1/2} \prec I $. Again, as $\etab$ is estimable, $\text{rank}(Z_c^TZ_c)=\text{rank}(Z^TZ)=r+c-1$. The last two inferences combined suggest that  $\lambda_j(Z^TZ) > \lambda_j(Z_c^TZ_c)$ for some $j \in \{1,\cdots, r+c-1\}$. By the Cauchy interlacing theorem, this is a contradiction as $Z$ was produced by deleting rows of $Z_c$, and so $Z^TZ$ is a compression of $Z_c^TZ_c$.


\section{Section~\ref{sec:computation} Details: URE Computations}\label{append.computation}

By definition, $\V = Z\Lambda \Lambda^{\topp} Z^{\topp} + M$. 
We apply the matrix inverse identity to get
\bel{eq:woodbury}
\V^{-1} = M^{-1} - M^{-1} Z \Lambda(\Lambda^{\topp} Z^{\topp} M^{-1} Z \Lambda + I_q)^{-1} \Lambda^{\topp} Z^{\topp} M^{-1}.
\eel
Hence, we have
\begin{align*}
M \V^{-1} &= I_{rc} - Z \Lambda(\Lambda^{\topp} Z^{\topp} M^{-1} Z \Lambda + I_q)^{-1} \Lambda^{\topp} Z^{\topp} M^{-1}\\
M \V^{-1} M &= M - Z \Lambda(\Lambda^{\topp} Z^{\topp} M^{-1} Z \Lambda + I_q)^{-1} \Lambda^{\topp} Z^{\topp}.
\end{align*}
Using the above, we get
\bel{eq:tr-vbar-inv-mbar-sq}
\begin{aligned}
	\tr(\V^{-1} M^2) &= \tr(M \V^{-1} M) 
	= \tr(M) - \tr( Z \Lambda ( \Lambda^{\topp} Z^{\topp} M^{-1} Z \Lambda + I_q )^{-1} \Lambda^{\topp} Z^{\topp} ).
\end{aligned}
\eel
Therefore, \eqref{eq:sure} can be written as
\bel{eq:sure-comp}
\URE = -\sigma^2 \tr(M) + 2 \sigma^2 \tr\{ (\Lambda^{\topp} Z^{\topp} M^{-1} Z \Lambda + I_q )^{-1} (\Lambda^{\topp} Z^{\topp} Z \Lambda) \} + \| M \V^{-1} (\y -\one\mu) \|^2.
\eel
In computing \eqref{eq:sure-comp}:
\benu
\item The middle term is computed as the sum of the \textit{elementwise} product of $(\Lambda^{\topp} Z^{\topp} M^{-1} Z \Lambda + I_q )^{-1}$ and $\Lambda^{\topp} Z^{\topp} Z \Lambda$, using the property $\tr(A^{\topp} B) = \sum_{i,j}A_{ij} B_{ij}$
\item $(\Lambda^{\topp} Z^{\topp} M^{-1} Z \Lambda + I_q )^{-1}$ is computed efficiently employing a sparse Cholesky factorization of $\Lambda^{\topp} Z^{\topp} M^{-1} Z \Lambda + I_q$ similarly to the implementation in the lme4 package in \texttt{R}. 
\item The quantity $\min_\mu \| M \V^{-1}(\y -\one\mu) \|^2$ is computed by regressing $M\V^{-1}\y$ on $M\V^{-1}\one_{rc}$ using the \texttt{lm} function in \texttt{R}. 
In doing that, the vector $M \V^{-1}\x$ (for $\x = \y$ and $\x =\one_{rc}$) is computed as:
\bel{eq:mbar-vbar-inv-x}
M \V^{-1} \x= \x - Z \Lambda ( \Lambda^{\topp} Z^{\topp} M^{-1} Z \Lambda + I_q )^{-1} \Lambda^{\topp} Z^{\topp} (M^{-1} \x)
\eel
where \eqref{eq:mbar-vbar-inv-x} is implemented proceeding ``from right to left" to always compute a product of a matrix and a \textit{vector}, instead of two matrices: 
First find $M^{-1} \x$, then find $(\Lambda^{\topp} Z^{\topp})(M^{-1} \x)$, and so on.
\eenu

\section{Supplementary Materials}
Detailed derivations and discussions of the results whose proofs were not provided in the main paper is presented here. Detailed discussions regarding the assumptions we made for the asymptotic theory are also provided here.
\subsection{Details and Proofs of results stated in Section~\ref{sec:bayes}}\label{append.sec2}

\begin{lemma}\label{lem.misc.1}
	Under model the hierarchical Gaussian model \eqref{eq:model-bayes}:\\
	(a) The marginal distribution of $\y$ is
	$$\y \sim N(\etab, \sigma^2 \V) \text{ where } \V = Z\Lambda \Lambda^{\topp} Z^{\topp} + M= \lambdaa \Za \Za^{\topp} + \lambdab \Zb \Zb^{\topp} + M.$$
	(b) The Bayes estimate of $\etab$ is 
	$$	\etahatb^{\sf Bayes} = \y - M \Vinv (\y - \mu \cdot \one) \;.$$
\end{lemma}

\noindent \textbf{Proof. } 
(a ) Write $\y = \one \mu + Z\btheta + \boldsymbol{\epsilon}$ where $\boldsymbol{\epsilon}\sim N_{rc}(\0,\sigma^2M)$ is independent of $\btheta$. 
Clearly, $\y$ is Gaussian. 
Also, $\E(\y) = \one \mu$ and, by independence of $\btheta$ and $\boldsymbol{\epsilon}$, 
$\text{cov}(\y) = \text{cov}(Z\btheta) + \text{cov}(\boldsymbol{\epsilon}) = \sigma^2(Z\Lambda\Lambda^{\topp} + M)= \sigma^2 \Sigma$. \\
(b) Using the representation of $\y$ in (a), $(\y, \btheta)^\topp$ is Gaussian because it is a linear transformation of $(\btheta, \boldsymbol{\epsilon})^\topp$. 
Then, since $\text{cov}(\btheta,\y) = \sigma^2\Lambda\Lambda^\topp Z^\topp$, 
$
\E[\btheta|\y] = \text{cov}(\btheta,\y) [\text{cov}(\y)]^{-1} (\y-\mu \one) = \Lambda\Lambda^\topp Z^\topp \Sigma^{-1} (\y-\mu \one). 
$
Hence,
$$
\begin{aligned}
\E[\etab|\y] = \E[\one \cdot \mu + Z\btheta|\y] &= \mu \one + Z\E[\btheta|\y] \\
&= \one \cdot \mu + (\Sigma-M)\Sigma^{-1}(\y-\one \cdot \mu) = \y - M\Sigma^{-1}(\y-\one \cdot \mu). 
\end{aligned}
$$

\begin{lemma}\label{lem.misc.2}
	An unbiased estimate of the risk $R_{r,c}(\etab, \etahatsb(\mu,\lambdaa,\lambdab))$ is
	$$\URE(\mu, \lambdaa, \lambdab) = \rcinv \big\{ \sigma^2 \tr(M) - 2\sigma^2 \tr (\Vinv M^2) + (\y - \one\mu)^t [\Vinv M^2 \Vinv] (\y - \one\mu) \big \}.$$
\end{lemma}

\noindent \textbf{Proof. }
This is immediate from the formula in \citep[p. 362]{berger1985statistical} after noticing that $\y|\etab \sim N_{rc}(\etab,\sigma^2M)$ and writing $\etahatsb(\mu,\lambdaa,\lambdab) = \y - \sigma^2M(\sigma^2\Sigma)^{-1}(\y- \mu\cdot \one)$.

\subsubsection{Estimating Equations for \eqref{eq:model-twoway}}\label{append.estimating.eqn}

\noindent \textbf{Estimating Equations for the ML method.}
ML estimates are computed based on the likelihood of $\y$ in the hierarchical model \eqref{eq:model-bayes}. 
Our derivation is similar to the analysis conducted in Chapter 6.3, 6.4, 6.8 and 6.12 of \citet{searle2001generalized}. Since $\y \sim N_{rc}(\one \mu, \sigma^2 \V)$, its density is given by:
\bel{eq:lik}
f(\y)= \frac{1}{(2\pi\sigma^2)^{rc/2} |\V|^{1/2}} \exp\left\{ -\frac{1}{2\sigma^2} (\y -\one\mu)^t \V^{-1} (\y -\one\mu)\right\}
\eel
and the corresponding log-likelihood is
\bel{eq:loglik}
l(\mu, \btheta) = -(rc)/2 \cdot \log(2\pi\sigma^2) - \frac{1}{2} \log |\V| - \frac{1}{2\sigma^2}(\y -\one\mu)^t \V^{-1} (\y -\one\mu)
\eel
Using chain rule, we have
\begin{equation} \label{eq:dmu-ml}
\frac{\partial l}{\partial{\mu}} \stackrel{\eqref{eq:quad}}{=} -\frac{1}{\sigma^2}(\y -\one\mu)^t \V^{-1} \frac{\partial \{\y -\one\mu\}}{\partial \mu} = 
\frac{1}{\sigma^2}(\y -\one\mu)^t \V^{-1} \one
\end{equation}
Also,
\begin{align}
\frac{\partial l}{\partial{\lambdaa^2}} &\stackrel{(\texttt{R}\ref{eq:logdet})}{=} 
- \frac{1}{2} \tr \left( \V^{-1} \frac{\partial \V}{\partial \lambdaa^2} \right) -\frac{1}{2\sigma^2} (\y -\one\mu)^t \left[ \frac{\partial \V^{-1}}{\partial \lambdaa^2} \right] (\y -\one\mu) \notag \\
&= -\frac{1}{2} \left\{ \tr \left( \V^{-1} \frac{\partial \V}{\partial \lambdaa^2} \right) + \frac{1}{\sigma^2}(\y -\one\mu)^t \left[ \frac{\partial \V^{-1}}{\partial \lambdaa^2} \right] (\y -\one\mu) \right\} \notag  \\
&\stackrel{(\texttt{R}\ref{eq:invmat})}{=} -\frac{1}{2} \left\{ \tr \left( \V^{-1} \frac{\partial \V}{\partial \lambdaa^2} \right) - \frac{1}{\sigma^2} (\y -\one\mu)^t \V^{-1} \left[ \frac{\partial \V}{\partial \lambdaa^2} \right] \V^{-1} (\y -\one\mu) \right\} \notag \\
&= -\frac{1}{2} \left\{  \tr \left( \V^{-1} \Za \Za^{\topp} \right) - \frac{1}{\sigma^2} (\y - \mu \one)^t \V^{-1} \Za \Za^{\topp} \V^{-1} (\y - \mu \one)  \right\} \label{eq:dthetaasq-ml}
\end{align}
where in the last equality  we use the fact that
\bel{eq:vbar}
\V = \lambdaa^2 \Za \Za^{\topp} + \lambdab^2 \Zb \Zb^{\topp} + \sigma^2M.
\eel

\noindent On equating to zero, we get from \eqref{eq:dmu-ml} that the optimal estimate of the location parameter is given by
\bel{eq:mu-ml}
\hat{\mu}_1 = \frac{\one^{\topp} \V^{-1}\y }{\one^{\topp} \V^{-1}1 },
\eel
the GLS estimate of $\mu$. If $\hat{\mu}_1 \notin [\hat{a}_{\tau},\hat{b}_{\tau}]$, $\hat{\mu}$ takes the nearest boundary value in the set. 
From \eqref{eq:dthetaasq-ml}, we get the estimating equations for $\lambdaa$,
\bel{eq:thetaasq-ml}
\tr \left( \V^{-1} \Za \Za^{\topp} \right) - \frac{1}{\sigma^2} (\y -\one\mu)^t \V^{-1} \Za \Za^{\topp} \V^{-1} (\y -\one\mu) = 0.
\eel
By symmetry, taking the partial derivative w.r.t. $\lambdab^2$ gives
\bel{eq:thetabsq-ml}
\tr \left( \V^{-1} \Zb \Zb^{\topp} \right) - \frac{1}{\sigma^2} (\y -\one\mu)^t \V^{-1} \Zb \Zb^{\topp} \V^{-1} (\y -\one\mu) = 0.
\eel
 If $\hat{\mu}_1 \in [\hat{a}_{\tau},\hat{b}_{\tau}]$, plugging \eqref{eq:mu-ml} into \eqref{eq:thetaasq-ml} and \eqref{eq:thetabsq-ml} gives the estimating equations for $\lambdaa^2$ and $\lambdab^2$ as
\begin{align} \label{eq:est-eqn-ml}
\tr \left( \Vhat^{-1} \Za \Za^{\topp} \right) - \frac{1}{\sigma^2} \y^{\topp}(I - P)^t \Vhat^{-1} \Za \Za^{\topp} \Vhat^{-1} (I - P) \y &= 0 \\
\tr \left( \Vhat^{-1} \Zb \Zb^{\topp} \right) - \frac{1}{\sigma^2} \y^{\topp}(I - P)^t \Vhat^{-1} \Zb \Zb^{\topp} \Vhat^{-1} (I - P) \y &= 0
\end{align}
where $P$ is the Generalized Least Square projection matrix:
\bel{eq:proj-ml}
P =\one (\one^{\topp} \Vhat^{-1}\one)^{-1}\one^{\topp} \Vhat^{-1}.
\eel

\noindent \textbf{Estimating Equations for the URE method.}\\[2ex]
For URE estimates, note that in \eqref{eq:sure}, in comparison to \eqref{eq:loglik}, $\V^{-1}M^2V^{-1}$ replaces $\V^{-1}$. 
Hence the partial derivative w.r.t. $\mu$ vanishes for 
\bel{eq:mu-sure-1}
\hat{\mu}_1 = \frac{\one^{\topp} [\V^{-1}M^2V^{-1}] y }{\one^{\topp} [\V^{-1}M^2V^{-1}]\one }.
\eel
Again, if $\hat{\mu}_1 \notin [\hat{a}_{\tau},\hat{b}_{\tau}]$ it takes the nearest boundary value of the set. 
Furthermore, 
\begin{align}
&\frac{\partial}{\partial{\lambdaa^2}} \URE = \stackrel{(\texttt{R}\ref{eq:prod})}{=}-2\sigma^2 \tr \left( \frac{\partial \V^{-1}}{\partial \lambdaa^2} M^2 \right) + (\y -\one\mu)^t \left\{ \frac{\partial \V^{-1}}{\partial \lambdaa^2} M^2 \V^{-1} + \V^{-1} M^2 \frac{\partial \V^{-1}}{\partial \lambdaa^2} \right\} (\y -\one\mu) \notag \\
&= -2\sigma^2 \tr \left( \frac{\partial \V^{-1}}{\partial \lambdaa^2} M^2 \right) + 2 (\y -\one\mu)^t \left[ \frac{\partial \V^{-1}}{\partial \lambdaa^2} M^2 \V^{-1} \right] (\y -\one\mu) \notag \\
&\stackrel{(\texttt{R}\ref{eq:invmat})}{=}2\sigma^2 \tr \left( \V^{-1} \frac{\partial \V}{\partial \lambdaa^2} \V^{-1} M^2 \right) - 2 (\y -\one\mu)^t \left[ \V^{-1} \frac{\partial \V}{\partial \lambdaa^2} \V^{-1} M^2 \V^{-1} \right] (\y -\one\mu) \notag \\
&= 2\sigma^2 \tr(\V^{-1} \Za \Za^{\topp} \V^{-1} M^2) - 2(\y -\one\mu)^t [\V^{-1} \Za \Za^{\topp} \V^{-1} M^2 \V^{-1}] (\y -\one\mu) \label{eq:dthetaasq-sure}
\end{align}
Hence, on equating \eqref{eq:dthetaasq-ml} to zero we obtain
\bel{eq:thetaasq-sure}
\tr(\V^{-1} \Za \Za^{\topp} \V^{-1} M^2) - \frac{1}{\sigma^2} (\y -\one\mu)^t [\V^{-1} \Za \Za^{\topp} \V^{-1} M^2 \V^{-1}] (\y -\one\mu) = 0
\eel
By symmetry, equating the partial derivative w.r.t. $\lambdab^2$ to zero gives
\bel{eq:thetabsq-sure}
\tr(\V^{-1} \Zb \Zb^{\topp} \V^{-1} M^2) - \frac{1}{\sigma^2} (\y -\one\mu)^t [\V^{-1} \Zb \Zb^{\topp} \V^{-1} M^2 \V^{-1}] (\y -\one\mu) = 0
\eel
If $\hat{\mu}_1 \in [\hat{a}_{\tau},\hat{b}_{\tau}]$, plugging \eqref{eq:mu-sure-1} into \eqref{eq:thetaasq-sure} and \eqref{eq:thetabsq-sure} gives the estimating equations for $\lambdaa^2, \lambdab^2$ as 
\begin{align} \label{eq:est-eqn-sure}
\tr \left( \Vhat^{-1} \Za \Za^{\topp} \Vhat^{-1} M^2 \right) - \frac{1}{\sigma^2} \y^{\topp}(I - P)^t \V^{-1} \Za \Za^{\topp} \Vhat^{-1} M^2 \Vhat^{-1} (I - P) \y &= 0 \\
\tr \left( \Vhat^{-1} \Zb \Zb^{\topp} \Vhat^{-1} M^2 \right) - \frac{1}{\sigma^2} \y^{\topp}(I - P)^t \V^{-1} \Zb \Zb^{\topp} \Vhat^{-1} M^2 \Vhat^{-1} (I - P) \y &= 0
\end{align}
where $P$ is given by:
\bel{eq:proj-sure}
P =\one (1^{\topp} \Vhat^{-1} M^2 \Vhat^{-1}1)^{-1}1^{\topp} \Vhat^{-1} M^2 \Vhat^{-1}.
\eel

\subsection{Section~\ref{sec:missing} Appendix}\label{append.missing}

\noindent \textbf{Proof of Lemma~\ref{lem.missing.1}.}
This is an immediate consequence of Theorem 5 in \citet{searle1966est}, because $\etatil = \Ztil\btheta$ is estimable if and only if $v^\topp \btheta$ is estimable for each row $v$ of $\Ztil$. 

\begin{lemma}\label{lem.missing.3}
	An unbiased estimator of the generalized risk $R^Q_{r,c}(\etab, \etahatsb(\mu,\lambdaa,\lambdab))$ of estimators of the form $\etahatsb(\mu,\lambdaa,\lambdab)$ is given by
	\begin{align*}
	\begin{split}
	\UREQ (\mu, \lambdaa, \lambdab)  =  \sigma^2 \tr(QM)  &- 2\sigma^2 \tr (\Vinv MQM) \\
	&+ (\y - \mu \one)^t \big[\Vinv MQM \Vinv\big] (\y -  \mu \one). 
	\end{split}
	\end{align*}
\end{lemma}

\noindent \textbf{Proof. } Similar to Lemma~\ref{lem.misc.2}.

\subsection{Section~\ref{sec:theory}: Supplementary Materials} 
\noindent \textbf{Proof of Lemma~\ref{lem:sec4.temp.1}.}
As $\hat{a}_{\tau}, \hat{b}_{\tau}$ is the $\tau/2$ th and $(1-\tau/2)$ th quantile of $\y$:
\begin{align*}
\max(|\hat{a}_{\tau}|,|\hat{b}_{\tau}|) & \leq \texttt{quantile} (|y_{ij}|: (i,j)\in \mathcal{E}; 1 -\tau/2)\\ 
& = \texttt{quantile}(|\eta_{ij}| + |\epsilon_{ij}|: (i,j) \in \mathcal{E}; 1 -\tau/2 )
\end{align*}
where $\epsilon_{ij}$ are i.i.d. standard normal variables. The RHS is bounded above by:
\begin{align*}
\max\, \big \{|\eta_{ij}| + |\epsilon_{ij}|: (i,j) \in \mathcal{E}\text{ and } |\eta_{ij}| \leq q_{\tau}(|\eta|) , |\epsilon_{ij}| \leq q_{\tau}(|\epsilon|) \big\}
\leq q_{\tau}(|\eta|) + q_{\tau}(|\epsilon|)~, 
\end{align*}
where $q_{\tau}(|\eta|)=\texttt{quantile}(|\eta_{ij}|: (i,j) \in \mathcal{E}; 1 -\tau/2)$ and $q_{\tau}(|\epsilon|)=\texttt{quantile}(|\epsilon_{ij}|: (i,j) \in \mathcal{E}, 1 -\tau/2)$. Thus, 
\begin{align*}
\max(|\hat{a}_{\tau}|,|\hat{b}_{\tau}|) \leq q_{\tau}(|\eta|) + q_{\tau}(|\epsilon|).
\end{align*}
Again,
$$ q_{\tau}(|\eta|) \leq \max \{1, \texttt{quantile}(\eta_{ij}^2: (i,j) \in \mathcal{E}; 1 -\tau/2)\} \leq \max \bigg \{1, \frac{1}{\tau/2 \cdot RC} \sum_{i,j} \eta_{ij}^2 \bigg\} < \infty$$
which follows from Assumption A1. The second inequality above is due to the fact that the highest possible value of the $1 -\tau/2$ quantile of a series of positive numbers with a constraint on their sum is attained when all the values above that quantile are all same.\\
Also, as sample quantiles are asympotically normally distributed we have:\\
$$(rc)^{1/2} \cdot \big(q_{\tau}(|\eta|) -x_0 \big) \sim N(0, 8^{-1}\tau (1-\tau/2) \phi^{-2}(x_0)) \text{ where } x_0= \Phi^{-1}(1-\tau/4).$$
Thus, we have $P(\max(|\hat{a}_{\tau}|,|\hat{b}_{\tau}|) \leq \log(rc)) \to 1 \text{ as } r,c \to \infty$. This, completes the proof of the lemma.\\[1ex]

\noindent \textbf{Proof of Lemma~\ref{lem:sec4.temp}}
With a slight abuse of notation, we use $L^Q\big(\etab, \etahatsb(\mu, \lambdaatil, \lambdabtil)\big)$ to denote the loss $L^Q\big(\etab, \etahatsb(\mu, \lambdaa, \lambdab)\big)=L^Q\big(\etab, \etahatsb(\mu, 1/\lambdaatil^2-1, 1/\lambdabtil^2-1)\big)$. As $L^Q\big(\etab, \etahatsb(\mu, \lambdaatil, \lambdabtil)\big)$ is everywhere differentiable, for any triplet $(\mu, \lambdaatil, \lambdabtil)$ and any point $(\mu[i], \lambdaatil[j], \lambdabtil[k])$ on the grid $\net$ we have:
\begin{align*}
& \big \vert L^Q\big(\etab, \etahatsb(\mu, \lambdaatil, \lambdabtil)\big) - L^Q\big(\etab, \etahatsb(\mu[i], \lambdaatil[j], \lambdabtil[k])\big) \big \vert \\
&\leq D^{[1]}_{r,c} \cdot \big |\mu-\mu[i]\big | + D^{[2]}_{r,c} \cdot \big |\lambdaatil-\lambdaatil[j]\big | + D^{[3]}_{r,c} \cdot \big |\lambdabtil-\lambdabtil[k]\big |, \hspace{5cm}
\end{align*}
where,
\begin{align}\label{eq:D}
& D^{[1]}_{r,c}(\etab,\y) = \sup_{|\mu| \leq \thr; \lambdaatil, \lambdabtil \in [0,1];} \bigg \vert \frac{\partial}{\partial \mu}\, L^Q\big(\etab, \etahatsb(\mu, \lambdaatil, \lambdabtil)\big) \bigg \vert~ , \\ 
& D^{[2]}_{r,c}(\etab,\y) = \sup_{|\mu| \leq \thr; \lambdaatil, \lambdabtil \in [0,1];} \bigg \vert \frac{\partial}{\partial \lambdaatil}\, L^Q\big(\etab, \etahatsb(\mu, \lambdaatil, \lambdabtil)\big) \bigg \vert~\text{ and,} \\
& D^{[3]}_{r,c}(\etab,\y) = \sup_{|\mu| \leq \thr; \lambdaatil, \lambdabtil \in [0,1];} \bigg \vert \frac{\partial}{\partial \lambdabtil}\, L^Q\big(\etab, \etahatsb(\mu, \lambdaatil, \lambdabtil)\big) \bigg \vert~.
\end{align}
Thus, based on the construction of the grid $\nett_{r,c}$ we have for any triplet $(\mu, \lambdaa, \lambdab) \in [-\thr,\thr]\otimes [0,\infty] \otimes [0,\infty]$: 
\begin{align}\label{eq:D1}
& \inf_{(\mu[i], \lambdaa[j], \lambdab[k]) \in \nett_{r,c}} \big \vert L^Q\big(\etab, \etahatsb(\mu, \lambdaa, \lambdab)\big) - L^Q\big(\etab, \etahatsb(\mu[i], \lambdaatil[j], \lambdabtil[k])\big) \big \vert \\
& \leq D^{[1]}_{r,c}(\etab,\y) \cdot \delta^{[1]}_{r,c}+ D^{[2]}_{r,c}(\etab,\y) \cdot \delta^{[2]}_{r,c} + D^{[3]}_{r,c}(\etab,\y) \cdot \delta^{[3]}_{r,c} = {D}_{r,c} (\etab,\y) \text{ (say)}.
\end{align}
Thus,  on the set $A_{r,c}(\Y)=\{[\hat{a}_{\tau}, \hat{b}_{\tau}] \subseteq [-\thr,\thr]\}$ we have:
\begin{align*}
&  \big \vert L^Q\big(\etab, \etahatsb(\mu^{\rm OD}, \lambdaa^{\rm OD}, \lambdab^{\rm OD})\big) - L^Q\big(\etab, \etahatsb(\mu^{\rm OL}, \lambdaa^{\rm OL}, \lambdab^{\rm OL})\big) \big \vert \leq {D}_{r,c}(\etab,\y)~.
\end{align*}
By the construction of $\nett_{r,c}$ as shown afterwards in Lemma~\ref{lem:sec4.temp.3} we have: $\ex[{D}_{r,c}(\etab,\y)\, I\{A_{r,c}(\Y)\}] \to 0 $  as $r,c \to \infty$ under assumptions A1-A2.  It implies by Markov's inequality that
$P( {D}_{r,c}(\etab,\y) > \epsilon \text{ and } A_{r,c}(\Y)) \to 0$ as $r, c \to \infty$. 
These coupled with Lemmas~\ref{lem:sec4.temp.1} and \ref{lem:sec4.temp.2} provide us the results \texttt{A} and \texttt{B} of the lemma.

Again, note that by definition \eqref{eq:UD}, on the set $A_{r,c}(\Y)$  we have: 
\begin{align*}
&\big \vert L^Q\big(\etab, \etahatsb(\mu^{\rm UD}, \lambdaa^{\rm UD}, \lambdab^{\rm UD})\big) - L^Q\big(\etab, \etahatsb(\mu^{\rm URE}, \lambdaa^{\rm URE}, \lambdab^{\rm URE})\big) \big \vert \\
&\leq D^{[1]}_{r,c}(\etab,\y) \cdot \delta^{[1]}_{r,c}+ D^{[2]}_{r,c}(\etab,\y) \cdot \delta^{[2]}_{r,c} + D^{[3]}_{r,c}(\etab,\y) \cdot \delta^{[3]}_{r,c} = {D}_{r,c}(\etab,\y).
\end{align*}
and so the results \texttt{C} and \texttt{D} of the lemma follows using  Lemma~\ref{lem:sec4.temp.1} and result \texttt{B} of Lemma~\ref{lem:sec4.temp.2}.

\begin{lemma}\label{lem:sec4.temp.3}
	With $D_{r,c}(\etab,\y)$ defined in \eqref{eq:D}-\eqref{eq:D1}, for any $\etab$ obeying assumption A1 and under assumption A2 on the design we have:
	$$\ex[{D}_{r,c}(\etab,\y)] \to 0 \text{ as } r, c \to \infty.$$
\end{lemma}

\noindent \textbf{Proof of Lemma~\ref{lem:sec4.temp.3}.}
First, note that the quadratic loss is
$$L^Q\big(\etab, \etahatsb(\mu, \lambdaa, \lambdab))=(rc)^{-1}(\etab -\y + G \y - \mu G \one)^TQ(\etab - \y + G \y - \mu G \one)~,$$
where $G=M\Vinv$ and $\V=(\lambdaa Z_AZ_A^{\topp} +\lambdab Z_BZ_B^{\topp} + M )$ involves the scale parameters. Differentiating the loss with respect to $\mu$ we have:
\begin{align*}
\frac{\partial}{\partial \mu}\, L^Q\big(\etab, \etahatsb(\mu, \lambdaa, \lambdab)\big) &= (rc)^{-1}\,\frac{\partial}{\partial \mu}\,\big\{ \mu^2 \one^{\topp} G^{\topp} Q G \one - 2 \mu\, \one^{\topp} G^{\topp} Q(\etab-\y+G\y) \big\}\\
&=(rc)^{-1} \big(2 \, \mu\, \one^{\topp} G^{\topp} Q G \one - 2 \, \one^{\topp} G^{\topp} Q (\etab - \y+G \y)\big).
\end{align*}
Note that, 
$$(rc)^{-1} |\mu| \, \one^{\topp} G^{\topp} Q G \one \leq \thr \lambda_1(H) \text{ where } H = G^{\topp} Q G $$
and by calculations in Section~\ref{append.theory} of the appendix it follows that $ \lambda_1(H) \leq \nu_{r,c} \lambda_1(Q)$ for any $\lambdaa, \lambdab \geq 0$.\\
Also,
$\one^{\topp} G^{\topp} Q (\etab - \y+G \y) \sim N(\one^{\topp} H \etab, \,\one^TG^{\topp} Q (I-G^{\topp})M(I-G)G)\one)$ and by moment calculations similar to  Section~\ref{append.theory}  we have:
$$(rc)^{-1} \ex\{|\one^{\topp} G^{\topp} Q (\etab - \y+G \y)| \} \leq O(\nu_{r,c} \lambda_1(Q)) \text{ for any } \lambdaa, \lambdab \geq 0.$$
Therefore, $D^{[1]}_{r,c}(\etab,\y) \leq O(\thr\,\nu_{r,c}\, \lambda_1(Q))$ and so, $\ex \{D^{[1]}_{r,c}(\etab,\y) \delta^{[1]}_{r,c}\} \to 0$ as $r, c \to \infty$.
\par
Now, we concentrate on the scale hyper-parameters. Differentiating the loss with respect to $\lambdaa$ we have:
$$\frac{\partial}{\partial \lambdaa}\, L^Q\big(\etab, \etahatsb(\mu, \lambdaa, \lambdab)\big)=(\y-\mu \one)^T \frac{\partial  (G^{\topp} Q G)}{\partial \lambdaa} (\y -\mu \one) + 2 (\y-\mu \one)^T \frac{\partial G^{\topp}}{\partial \lambdaa} Q (\etab -\y)~,$$
$$ \text{ where, } \frac{\partial}{\partial \lambdaa} (G^{\topp} Q G)=\frac{\partial G^{\topp}}{\partial \lambdaa} Q G + G^{\topp} Q \frac{\partial G}{\partial \lambdaa} \quad \text{ and } \quad \frac{\partial G}{\partial \lambdaa}= M \Vinv Z_AZ_A^{\topp} \Vinv. $$
Again, note that for the transformed scale hyper-parameter $\lambdaatil$:
\begin{align*}
\frac{\partial}{\partial \lambdaatil}\, L^Q\big(\etab, \etahatsb(\mu, \lambdaatil, \lambdabtil)\big)&=\frac{\partial}{\partial \lambdaa}\, L^Q\big(\etab, \etahatsb(\mu, \lambdaa, \lambdab)\big) \times \frac{\partial \lambdaa}{\partial \lambdaatil}\\
&=-2 (1+\lambdaa)^{3/2}\frac{\partial}{\partial \lambdaa}\, L^Q\big(\etab, \etahatsb(\mu, \lambdaa, \lambdab)\big)~.
\end{align*}
Note that the change of scale to $\lambdatil$ was chosen cleverly such that not only the range of $\lambdatil$ is bounded but also the subsequent change in scale does not lead to the derivative to blow up as $\lambdaa$ varies over $0$ to $\infty$. As such:   
\begin{align*}
-\frac{1}{2}M^{-1}\frac{\partial G}{\partial \lambdaatil} &=  (1+\lambdaa)^{3/2} \Vinv Z_AZ_A^{\topp} \Vinv \\
&\preceq  \bigg(\lambdaa (1+\lambdaa)^{-3/4} Z_AZ_A^{\topp} +\lambdab(1+\lambdaa)^{-3/4} Z_B Z_B^{\topp} + (1+\lambdaa)^{-3/4} M \bigg)^{-1}. 
\end{align*}
As $\lambdaa \to \infty$, $(1+\lambdaa)^{-3/4} M$ becomes negligible  but  $\lambdaa (1+\lambdaa)^{-3/4} Z_AZ_A^{\topp}$  contributes massively and using moment calculations similar to Section~\ref{append.theory} of the Appendix, it can be shown that:$\ex \{D^{[2]}_{r,c}(\etab,\y)\} \leq O(\thr^2 \nu_{r,c} \, \lambda_1(Q))$. Similar, calculations hold for the other scale hyper-parameter. Combining the bounds on the three hyper-parameters, we get: $\ex[{D}_{r,c}(\etab,\y)] \to 0 \text{ as } r, c \to \infty.$\\[1ex]
\noindent \textbf{Proof of Lemma~\ref{lem:sec4.temp.final}} The proof is very similar to that of Lemma~\ref{lem:sec4.temp} and is avoided here to prevent repetition.\\[1ex]
\noindent \textbf{Proof of Lemma~\ref{lem:sec4.temp.2}}
To prove the $L_1$ convergence results of the lemma,  we apply Cauchy-Schwarz inequality and convert our problem to showing convergence of the products of the respective expected values. As such, 
\begin{align*}
\ex \big\{ | L_{r, c} (\etatil, \etatb_{\sf c}^{\rm OD}) -  L_{r, c} (\etatil, \etatb_{\sf c}^{\rm OL})| \cdot I\{A_{r,c}(\Y)\} \big\} &\leq 2 \ex \big\{ | L_{r, c} (\etatil, \etatb_{\sf c}^{\rm OD})| I\{A_{r,c}(\Y)\} \big\}\\
&\leq 2 \big \{ \ex \big\{ L^Q_{r, c} (\etab, \etatb^{\rm OD})\}^2 P(A_{r,c}(\Y)) \big\}^{1/2}.
\end{align*}
Based on the calculations made in the proof of Lemma~\ref{lem:sec4.temp.1}, it follows that $P(A_{r,c}(\Y))=O((rc)^{-1})$. Using moment bounding techniques used in Section~\ref{append.theory}, under assumptions A1 and A2, it can be shown that
$(rc)^{-1} \ex \big\{ L^Q_{r, c} (\etab, \etatb^{\rm OD})\}^2$, $ (rc)^{-1} \ex \big\{ L^Q_{r, c} (\etab, \etahatb^{\rm URE})\}^2$ and $(rc)^{-1}\ex \big\{ L^Q_{r, c} (\etab, \etahatb^{\rm UD})\}^2$ all converges to $0$ as $r,c \to \infty$ which will complete the proof of the lemma. \\[1ex]

\subsubsection{Brief Outline of the results for the Weighted loss case}\label{append.weighted}
We now briefly discuss estimation under weighted loss $L_{r,c}^{\textsf{wgt}}(\etab, \etahatb)$ defined in Section~\ref{sec:bayes}. For simplicity, we describe the case where there are no unobserved cells. Under this weighted loss, applying the following linear transformation 
\bes
\ytil = M^{-1/2}\y,\ \ \ \ \ \tilde{\etab} = M^{-1/2}\etab, \ \ \ \ \ \tilde{Z} = M^{-1/2}Z, \ \ \ \ \ \mutil \one = M^{-1/2}  \mu \one
\ees
the problem reduces to estimating $\tilde{\etab}$ from $\ytil\sim N(\tilde{\etab}, \sigma^2I)$ under the usual sum-of-squares loss. As the problem can be converted into a homoskedastic case, estimation here is easier than the cases discussed before. Assuming the hierarchical Gaussian prior structure like before, the complete Bayes model is given by:
\bes
\tilde{\etab} \sim N_{rc}(1\mutil, \sigma^2 M^{-1/2} Z\Lambda \Lambda^{\topp} Z^tM^{-1/2})
\ees
and the corresponding Bayes estimate of $\tilde{\etab}$ is
\bes
\hat{\etab} = \ytil - \Vtil^{-1}(\ytil - 1\mutil), \text{ where } \Vtil = M^{-1/2} Z\Lambda \Lambda^{\topp} Z^tM^{-1/2} + I
\ees
which unlike the shrinkage matrix in \eqref{eq:bayes} is symmetric. The oracle optimality proof can be worked out following in verbatim the proofs with the $L^Q$ loss.  However, in this case due to the presence of symmetric shrinkage matrix, the estimation problem reduces to the easier situation when $\nu_{r,c}=1$. 

\subsubsection{Discussions on the relevance of the Assumptions made}\label{append.assumption}
Here, we  discuss the genesis of Assumption A2 in our asymptotic optimality proofs. Our Assumption A1 is not very restrictive and so discussions on it is avoided here.  On the other hand, assumption A2 put an asymptotic control on the imbalance in our design matrix as $r, c \to \infty$. It is peculiar to the two-way nature of the problem and was never seen in the huge literature around shrinkage estimation of the normal mean in the one-way problem. 
\par
Assumption~A2 is needed in several parts of our proof. Let us concentrate on Lemma~\ref{lem:risk-1} which shows that our risk estimation strategy indeed approximates the true risk uniformly well for estimators with the location hyper-parameter $\mu$ set at $0$. By equation \eqref{eq:var-sure-zero}, the approximation error was exacted evaluated to be: 
$$ \ex \Big\{ \UREQ_{r, c}( 0, \lambdaa, \lambdab) - R^Q_{r,c}(\etab, \etahatsb(0, \lambdaa, \lambdab)) \Big\}^2 = (rc)^{-2} \{ 2 \tr(HMHM) + 4\etab^tHMH\etab \}.$$
We need to show that the RHS is $o(d_{r,c}^2)$ uniformly over any choices of the scale hyper-parameters and for all $\etab$ satisfying Assumption A1. 
Recall, $d_{r,c}^2$ rate of control of the square error was needed due to the discretization process. 
We concentrate on the component  $\etab^tHMH\etab$. Based on the equality condition on the von-Neumann trace inequality we can say that
$$\etab^tHMH\etab=\tr(\{HMH\} \{\etab\etab^{\topp}\}) = \lambda_1(HMH)  \etab^{\topp}\etab$$
when the eigen vector corresponding to the largest eigen value of $HMH$ matches $\etab/\etab^{\topp}\etab$. 
This, can indeed happen as for uniform convergence we not only have to consider all possible values $\etab$ but also all possible values of the $H$ matrix as $\lambdaa$, $\lambdab$ changes. To simplify further let us assume $Q=I$. We now provide heuristic reasons why $\lambda_1(HMH)$ can be close to the upper bound $\lambda_1^{-1}(M)$ that we use for it in our proofs.  As shown before:
$$\lambda_1(HMH)= \sigma^2_1(M^{-1/2}WM).$$
Now, $M$ is a diagonal matrix with $0 \prec M \preceq I$ and $0 \preceq W \preceq I$. $W$ depends on $\lambdaa$, $\lambdab$ as they vary over $[0,\infty]^2$. We relax the range and consider $M$ and $W$ to be any possible p.d. diagonal matrix and n.n.d. matrix respectively. It is difficult to gauge the degree of this tightness of the relaxation as $M$ and $W$ are related, but we can expect them to be close as $\lambdaa$ and $\lambdab$ span over the entire first quadrant. Simplifying the scenario further assume a $2\times 2$ situation where
$$ M= \left[ {\begin{array}{cc} 1 & 0 \\ 0 & b \end{array} } \right] \qquad \text{ and } \qquad  W= \left[ {\begin{array}{cc} w_{11} & w_{12} \\ w_{12} & w_{22} \end{array} } \right]~. $$
where $b \in (0,1]$ and $w_{11}, w_{12}$ and $w_{22}$ are chosen such that $0 \preceq W \preceq I$.
Thus, $(M^{-1/2}WM)(M^{-1/2}WM)^t$ is given by:
$$ \left[ {\begin{array}{ll} c_{11}=w_{11}^2+b^4 a_{12}^2 \qquad & c_{12}=b^{-1} w_{11}w_{21} + b^3 w_{22} w_{12} \\  c_{12} & c_{22}=b^{-2} w_{12}^2+b^2 w_{22}^2 \end{array} } \right] $$
and its  eigenvalues are given by:
$$2^{-1}\big\{(c_{11}+c_{22}) \pm \sqrt{(c_{11}+c_{22})^2+4c_{12}^2} \; \big\}.$$
We would like to evaluate the maximum of the eigenvalue as $b$ decreases.  We consider finding the eigenvalue asymptotically as $b \to 0$. 
Under the asymptotic regime $b\to 0$, we have:
$$c_{11}\sim w_{11}^2; \quad  c_{12} \sim b^{-1} w_{11}w_{21}, \quad \text{ and } \quad c_{22} \sim b^{-2} w_{12}^2~~. $$
Thus, for any fixed positive value of $w_{11}, w_{12}$ the highest eigenvalue is of the order of $b^{-1}=\lambda_1(M^{-1})$ as $b$ approaches zero.  
%

\subsection{Section~\ref{sec:bal} details: URE in Balanced Designs}\label{append.bal}
\noindent \textbf{Details of the risk decomposition in \eqref{eq:lin}}
This risk of the Bayes estimator $\etahatsb(y_{\cdot \cdot}, \lambdaa, \lambdab)$ in the balanced case is given by
\begin{align}
R(\etab,& \etahatsb(y_{\cdot \cdot}, \lambdaa, \lambdab)) = \notag \\
&= \frac{1}{rc} \ex \Big\{  \sum_{i=1}^r \sum_{j=1}^c [   (\mhatls - m) +  (c_{\alpha} \ahatls_i - a_i) +  (c_{\beta} \bhatls_j - b_j) ]^2  \Big\} \label{eq:risk-bal} \\
&= \frac{1}{rc} \ex \Big\{  rc(\mhatls - m)^2 + c \sum_{i=1}^r(c_{\alpha} \ahatls_i - a_i)^2 + r \sum_{j=1}^c(c_{\beta} \bhatls_j - b_j)^2  \Big\} \label{eq:orth} \\
&= \ex\Big\{  (\mhatls - m)^2  \Big\} + \frac{1}{r} \ex\Big\{  \sum_{i=1}^r(c_{\alpha} \ahatls_i - a_i)^2 \Big\} + \frac{1}{c} \ex\Big\{  \sum_{j=1}^c(c_{\beta} \bhatls_j - b_j)^2 \Big\} \label{eq:lin}
\end{align}
where equality \eqref{eq:orth} is due to orthogonality of the vectors corresponding to the three sums-of-squares. Note that that independence of $\mhatls, \ahatls, \bhatls$, which holds in the balanced case, was not needed in \eqref{eq:risk-bal}-\eqref{eq:ls-marginal}. 
Specifically, \eqref{eq:orth} holds also for unbalanced design because of the side conditions satisfied by $a,b$ and $\ahatls, \bhatls$; 
and \eqref{eq:ls-marginal} holds, with some known covariance matrices, in general for the generalized least squares estimators. 
Hence the calculation goes through for unbalanced data as well. 
However, in the unbalanced case \eqref{eq:sure-bs} no longer holds, i.e., the Bayes estimates for $\ba$ and $\bb$ are each functions of both $\ahatbls$ and $\bhatbls$.

\subsection{A list of some basic results used in our proofs} \label{sec:lin_results}

The following basic matrix algebra results are used in our proofs:
\benu[\texttt{R}1.]
\item For p.s.d. matrices $A, B$, if $0 \prec B\preceq A$, then $A^{-1} \preceq B^{-1}$ and $\lambda_k(B) \leq \lambda_k(A)$ for any $k$.\label{fc:psd-inv}
\item For p.s.d matrices $A, B$, $BAB$ is also p.s.d.  \label{fc:psd-butterfly}
\item For p.s.d matrices $A, B$, $\lambda_k(AB)\leq \lambda_k(A) \cdot \lambda_k(B)$ for any $k$.\label{fc:psd-eigen-prod}
\item For any matrices $C$ and $D$, $\sigma_1(CD) \leq \sigma_1(C) \cdot \sigma_1(D)$. \label{fc:singular-prod} 
\item (Von Neumann Trace inequality) If $C$ and $D$ are $n \times n$ Hermitian matrices then:
$$ \sum_{i=1}^n \lambda_i(A) \lambda_{n-i+1}(B) \leq \tr(AB) \leq \sum_{i=1}^n \lambda_i(A) \lambda_i(B).$$
Equality holds on the right when $B=\sum_{i=1}^n \lambda_i(B) u_i U_i^*$, and equality holds  on the left when $B=\sum_{i=1}^n \lambda_{n-i+1}(B) u_i U_i^*$ where $u_i$ is the right eigenvector of A for the eigen value $\lambda_i(A), i=1,\ldots,n$.
\label{fc:von-neumann}
\item For any matrix $C$,  $\sigma_1(C^{\topp}C) = \sigma_1(CC^{\topp})$ \label{fc:interchange}\\[3ex]
\noindent The following facts about derivatives involving matrix expressions are used in our paper. For matrices $U, B$ and $V$ where $B$ is independent of $x$ we have:
\item $\frac{\partial}{\partial x} \{ x^{\topp} B x \} = x^{\topp}(B + B^{\topp})$ \label{eq:quad}
\item $\frac{\partial}{\partial x} \log |A| = \tr( A^{-1} \frac{\partial A}{\partial x})$ \label{eq:logdet}
\item $\frac{\partial}{\partial x} A^{-1} = -A^{-1}\frac{\partial A}{\partial x} A^{-1}$ \label{eq:invmat}
\item $\frac{\partial}{\partial x} \{U B V\} = \frac{\partial U}{\partial x} BV + UB\frac{\partial V}{\partial x}$ \label{eq:prod} 
\end{enumerate}

\bibliographystyle{natbib}
\bibliography{mybib}

\end{document}